\documentclass[10pt]{article}
\usepackage[dvips]{graphics,color}
\usepackage{mathrsfs}
\usepackage{amssymb,amsmath,amsthm,mathrsfs,bm,subfigure,verbatim,appendix}
\usepackage{lastpage}
\usepackage{fancyhdr}%
\usepackage{array, multirow}
\usepackage{listings}
\usepackage{natbib}
\usepackage{graphicx}
\usepackage{mathrsfs}
\usepackage{multirow}

\theoremstyle{plain}
\newtheorem{theorem}{Theorem}[section]
\theoremstyle{plain}

\theoremstyle{plain}

\theoremstyle{remark}

\theoremstyle{plain}
\newtheorem{condition}{Condition}

\begin{document}

\markboth{S. Kong and B. Nan }{Covariate subject to limit of
detection}

\title{Semiparametric Approach for
Regression with Covariate Subject to Limit of Detection}

\author{Shengchun Kong and Bin Nan}

\maketitle

\begin{abstract}
We consider generalized linear regression analysis with
left-censored covariate due to the lower limit of detection.
Complete case analysis by eliminating observations with values below
limit of detection yields valid estimates for regression
coefficients, but loses efficiency; substitution methods are biased;
maximum likelihood method relies on parametric models for the
unobservable tail probability distribution of such covariate, thus
may suffer from model misspecification. To obtain robust and more
efficient results, we propose a semiparametric likelihood-based
approach for the estimation of regression parameters using an
accelerated failure time model for the covariate subject to limit of
detection. A two-stage estimation procedure is considered, where the
conditional distribution of the covariate with limit of detection
given other variables is estimated prior to maximizing the
likelihood function. The proposed method outperforms the complete
case analysis and the substitution methods as well in simulation
studies. Technical conditions for desirable asymptotic properties
are provided.
\end{abstract}


\section{Introduction}

Detection limit is a threshold below which measured values are not
considered significantly different from background noise
\citep{helsel:2005}. Hence, values measured below this threshold are
unreliable. In environmental epidemiology, particularly exposure
analysis, when exposure levels are low, measurement of chemicals has
a large percentage falling below the limit of detection due to
inadequate instrument sensitivity. This is a common issue in the
National Health and Nutrition Examination Survey
\citep{Crainiceanu:2008}, where many exposure variables have large
proportions of measurements below their limits of detection. For
example, 27.8\% of the urine arsenobetaine measures are below its
limit of detection at 0.4 $\mu$g/l \citep{Caldwell2009}. In the
Diabetes Prevention Program, 66 of the 301 eligible participants had
their testosterone levels below the detection limit of 8.0 ng/dl
\citep{Kim:2012}. In an analysis for the Michigan Bone Health and
Metabolism Study, up to 66\% of the 50 study participants had
anti-Mullerian hormone below the limit of detection at 0.05ng/ml
\citep{Sowers:2008}. For illustrative purpose in this article, we
consider an analysis for the National Health and Nutrition
Examination Survey, which examines the relationship between arsenic
exposure and the prevalence of type 2 diabetes
\citep{Navas-Acien:2008}.

A variable with limit of detection can be either a response variable
or a covariate in regression analysis. We focus on the latter in
this article. Although many ad hoc methods have been implemented in
practice, more appropriate statistical methods for regression models
with a covariate subject to limit of detection are yet to be
thoroughly studied \citep{Schisterman:2010}. The complete case
analysis, simply eliminating observations with values below limit of
detection, yields consistent estimates of the regression
coefficients \citep{nie:2010, little rubin:2002}, but loses
efficiency. Substitution methods are frequently used in epidemiology
studies, where the values of covariate $Z$ below limit of detection,
denoted by $L$, are substituted by $L$, or $L/\sqrt{2}$, or zero,
see for example, \cite{Schisterman:2006}, \cite{hornung:1990},
\cite{moulton:2002}, \cite{Koru-Sengul2011}, \cite{Kroger2009},
\cite{Boomsma 2009}, \cite{Bloom2008}, and \cite{Gollenberg2010}
among many others. These methods are easily implementable, but found
to be inappropriate and can yield large biases \citep{helsel:2006,
nie:2010}. \cite{Richardson and Ciampi:2003} proposed to replace the
values below limit of detection $L$ with $E(Z|Z < L)$, which is
obtained from an assumed known distribution of $Z$. There are two
issues with this approach, however, (i) the distributional
assumption is not verifiable; (ii) even if $E(Z|Z < L)$ is correctly
specified, the method only leads to consistent estimates in linear
regression when the covariate subject to limit of detection $Z$ is
independent of all the other covariates.

Another widely used method is the maximum likelihood estimation
based on a parametric distributional assumption to the unobservable
tail probability of the covariate that is subject to limit of
detection. For examples,  \cite{nie:2010} and \cite{Arunajadai 2012}
considered the linear regression based on a normal and a generalized
gamma distribution for the covariate subject to limit of detection,
respectively; \cite{Cole et al:2009}, \cite{wu2012} and \cite{Albert
2010} considered parametric maximum likelihood approach under the
generalized linear regression; \cite{D'Angelo and Weissfeld:2008}
applied this approach to the Cox regression. In practice, however,
the underlying covariate distribution is unknown. The test of the
parametric assumption to the unobservable tail probability of the
covariate that is subject to limit of detection is usually
unavailable because there is no information/observation below the
detection limit. Both \cite{lynn:2001} and \cite{nie:2010} noted
that a parametric assumption can yield large bias if misspecified
and argued that such an approach should not be attempted.
\cite{nie:2010} recommended the complete case analysis despite the
fact that simply dropping data below the limit of detection can lose
a significant amount of information.

To obtain more efficient and yet robust results, we propose a
semiparametric likelihood-based approach to fit generalized linear
models with covariate subject to limit of detection. The tail
distribution of the covariate beyond its limit of detection is
estimated from a semiparametric accelerated failure model,
conditional on all the fully observed covariates. Model checking can
be done using martingale residuals for semiparametric accelerated
failure time models. The proposed method is shown to be consistent
and asymptotically normal, and outperforms existing methods in
simulations. The proof of the asymptotic properties relies heavily
on empirical process theory, which is provided in the online
Supplementary Material.

\section{A semiparametric approach}
For a single observation, denote the response variable by $Y$, the
covariate subject to limit of detection by $Z$, and the fully
observed covariates  by $X=(X_{1}, \dots, X_{p})'$, where $p$ is the
number of fully observed covariates. For simplicity, we only
consider one covariate that is subject to limit of detection.
Consider a generalized linear model with
\begin{equation}
E(Y)=\mu=g^{-1}(D'\theta), \label{modellod1}
\end{equation}
where $g$ is the link function, $D'\theta$ is the linear predictor
with $D=(1,X',Z)'$ and $\theta=(\beta',\gamma)'$, here $\beta$ is a
$(p+1)$-dimensional vector and $\gamma$ is a scalar. The variance of
$Y$, typically a function of the mean, is denoted by
\begin{equation*}
var(Y)=W(\mu)=W\{g^{-1}(D'\theta)\}.
\end{equation*}
We consider the exponential dispersion family in the natural form
\citep{agresti:2002, mccullagh:1989} given $(Z,X)$
\begin{equation}
f_{\varpi,\phi}(Y|Z,X)=\exp\bigg\{\frac{Y\varpi-b(\varpi)}{a(\phi)}+c(Y,\phi)\bigg\},\label{glm}
\end{equation}
where $\phi$ is the dispersion parameter and $\varpi$ is the natural
parameter. We have $\mu=E(Y)=\dot{b}(\varpi)$, and
$var(Y)=\ddot{b}(\varpi)a(\phi)$, where $\dot{b}$ is the first
derivative of $b$ and $\ddot{b}$ is the second derivative of $b$.

The actual value of $Z$ is not observable when $Z<L$, where the
constant $L$ denotes the limit of detection, which is an example of
left-censoring. In practice $Z$ is a concentration measure of
certain substance and thus non-negative. Consider a monotone
decreasing transformation $h$ that yields $Z=h(T)$, for example,
$h(T)=\exp(-T)$. Denote $D(T)=(1,X',h(T))'$. If $T\leq C=h^{-1}(L)$,
then $T$ is observed; otherwise $T$ is right-censored by $C$. We
denote the observed value by $V=\min(T,C)$ and the censoring
indicator by $\Delta=I(T\leq C)$.

The proposed methodology works for a broad family of link functions
defined by the regularity conditions given in the Appendix. For
notational simplicity, we present the main material using canonical
link function $g$, where $g=(\dot{b})^{-1}$. Then, when $T$ is
observed, model (\ref{glm}) becomes
\begin{equation}
f_{\theta,\phi}(Y|T,X)=\exp\bigg\{\frac{YD'(T)\theta-b(D'(T)\theta)}{a(\phi)}+c(Y,\phi)\bigg\}.\label{modellod2}
\end{equation}

Denote the conditional cumulative distribution function of $T$ given
$X$ by $F_1(t|X)$ with density $f_1(t|X)$. The likelihood function
for the observed data $(V,\Delta,Y,X)$ can be factorized into
\begin{equation*}
f(V,\Delta,Y,X)=f_2(V,\Delta |  Y,X)f_3(Y |
X)f_4(X),\label{lodfact1}
\end{equation*}
where $f$ denotes the joint density of $(V, \Delta, Y, X)$, $f_2$
denotes conditional density of $(V,\Delta)$ given $(Y,X)$, $f_3$
denotes conditional density of $Y$ given $X$, and $f_4$ denotes
marginal density of $X$. Going through conditional arguments using
the Bayes' rule and dropping $f_4(X)$, we obtain the likelihood
function
\begin{equation}
L(V,\Delta,Y,X)=\{f_{\theta,\phi}(Y |  T,X)f_1(T |
X)\}^{\Delta}\left\{\int_C^{\infty} f_{\theta,\phi}(Y | t,X)dF_1(t |
X)\right\}^{1-\Delta},\label{lodfact2}
\end{equation}
where only $f_{\theta,\phi}$ contains the parameter of interest
$\theta$, whereas $f_1$ is a nuisance parameter in addition to
$\phi$.

There are two parts in (\ref{lodfact2}): (i) $\{f_{\theta,\phi}(Y |
T,X)f_1(T |  X)\}^{\Delta}$ for fully observed subject, and (ii)
$\left\{\int_C^{\infty} f_{\theta,\phi}(Y | t,X)dF_1(t |
X)\right\}^{1-\Delta}$ for subject with covariate below limit of
detection. Complete case analysis is only based on the first part
and, although it yields a consistent estimate of $\theta$, it
clearly loses efficiency.  We see from the second part of
(\ref{lodfact2}) that the efficiency gain comparing to the complete
case analysis depends on how well we can recover the right tail of
the conditional distribution $F_1(t |  X)$ beyond $C$. Parametric
models for $F_1(t |  X)$ are often considered in the literature, see
\cite{nie:2010}, but it may suffer from model misspecification. The
nonparametric method degenerates to the complete case analysis
because there is no actual observation beyond censoring time $C$. We
consider a semiparametric approach that allows reliable
extrapolation beyond $C$ and is robust against any parametric
assumption.

Among all the commonly used semiparametric models for right-censored
data, only the accelerated failure time model allows extrapolation
beyond $C$, and model checking can be done by visualizing the
cumulative sums of the martingale-based residuals \citep{lin:1993,
lin:1996, peng:2006}. We hence propose a semiparametric accelerated
failure time model for the transformed covariate subject to limit of
detection given by
\begin{equation}
T=X'\alpha+\varsigma,  \label{lodAFT}
\end{equation}
where $\varsigma$ follows some unknown distribution, denoted by
$\eta$, and is independent of $X$. We only consider a fixed $h$ for
$T$ in this article. More flexible transformation, for example, the
Box-Cox transformation \citep{BoxCox:1964, Foster:2001, Cai:2005},
is worth further investigation.  Note that $X$ appears in both
models (\ref{modellod1}) and (\ref{lodAFT}), but it may refer to
different forms of covariates in these models. For example, $X_1$ is
a covariate in (\ref{modellod1}) whereas $X_1^2$ is a covariate in
(\ref{lodAFT}). We use the same $X$ to denote all fully observed
covariates for notational simplicity. The log-likelihood function
then becomes
\begin{eqnarray}
\log L&=&\Delta \log f_{\theta,\phi}(Y |  T,X) + \Delta \log \dot{\eta}(T-X'\alpha) \nonumber \\
&& \qquad + (1-\Delta) \log \left\{\int_{C-X'\alpha}^{\tau}
f_{\theta,\phi}(Y | t+X'\alpha,X)d\eta(t)\right\},\label{lodlike1}
\end{eqnarray}
where $\tau$ is a truncation time at the residual scale defined in
Condition 4 in the Appendix.

\section{The pseudo-likelihood method}\label{lod:pseudo}

The log likelihood function (\ref{lodlike1}) involves an unknown
distribution function $\eta$ and its derivative, hence a
semiparametric maximum likelihood estimation, if it exists, can be
complicated. We propose a tractable two-stage pseudo-likelihood
approach in which the nuisance parameters $(\phi,\alpha,\eta)$ are
estimated in stage 1, and the parameter of interest $\theta$ is then
estimated by maximizing the data version of (\ref{lodlike1}) in
stage 2 with nuisance parameters replaced by their estimators
obtained in stage 1. Details are given below:

\medskip

{\em Stage} 1. Nuisance parameter estimation. Dispersion parameter
$\phi$ is estimated by the complete case analysis of the generalized
linear model (\ref{glm}); the accelerated failure time model
regression coefficient $\alpha$ is
 estimated by either the rank based methods, see \cite{wei:1990}, \cite{jin:2003},
 \cite{nan:2009} or the sieve maximum likelihood method, see \cite{ding:2011};
 and the accelerated failure time model error distribution $\eta$ is estimated by the
Kaplan-Meier estimator from the censored residuals.

The complete case analysis can be done by any standard statistical
package for generalized linear models. The rank based estimates for
the accelerated failure time model usually are obtained by using
linear programming. The R package ``rankreg" \citep{Zhou:2006}, now
archived by CRAN, can be implemented for small to moderate sample
sizes because it solves a linear programming problem of size $n^2$.
This is the method we implemented in simulations and the arsenic
exposure data example. An alternative approach is to modify the
Newton algorithm for solving the discrete rank based estimating
equation \citep{Yu and Nan:2006}. Standard Newton-Raphson algorithm
can be implemented to obtain the sieve maximum likelihood estimates
\citep{ding:2011} for the accelerated failure time model when the
sample size is large.

\medskip

{\em Stage} 2. Pseudo-likelihood estimation of $\theta$. Replacing
$(\phi,\alpha,\eta)$ by their Stage 1 estimates $(\hat\phi_n,
\hat\alpha_n, \hat\eta_{n, \hat\alpha_n})$ in the log likelihood
function yields the following log pseudo-likelihood function for a
random sample of $n$ observations:
\begin{eqnarray}
pl_n(\theta)&=&\frac{1}{n}\sum_{i=1}^n \bigg\{\Delta_i \log
f_{\theta,\hat{\phi}_n}(Y_i |  X_i,T_i)\nonumber\\
&& \qquad +(1-\Delta_i )\log \int_{C-X_i'\hat{\alpha}_{n}}^\tau
f_{\theta,\hat{\phi}_n}(Y_i |  X_i,t+X_i'\hat{\alpha}_{n})
d\hat{\eta}_{n,\hat{\alpha}_n}(t)\bigg\}, \label{lodpl}
\end{eqnarray}
where $$f_{\theta,\hat{\phi}_n}(Y_i |  T_i,X_i)=\exp\bigg[\frac{Y_i
\{D_i'(T_i)\theta\}-b\{D_i'(T_i)\theta\}}{a(\hat{\phi}_n)}+c(Y_i,\hat{\phi}_n)\bigg].$$
Note that the term $\Delta \log \dot\eta(T)$ in (\ref{lodlike1}) is
dropped because it does not involve $\theta$. We maximize
(\ref{lodpl}) by setting its derivative to zero and then solving the
equation to obtain the pseudo-likelihood estimator $\hat\theta_n$.
This is implemented by using standard Newton-Raphson algorithm with
the initial value obtained from the complete case analysis in Stage
1.

\medskip

Since $\hat\theta_n$ is obtained by solving an estimating equation,
its asymptotic properties can be obtained from Z-estimation theory.
It can be shown that all the estimates obtained in Stage 1 have
desirable statistical properties for Stage 2 estimation.  In
particular, $\hat\phi_n$ obtained from the complete case analysis is
$n^{1/2}$-consistent by \cite{little rubin:2002}; $\hat\alpha_n$ is
$n^{1/2}$-consistent by \cite{nan:2009} or \cite{ding:2011}; and
$\hat\eta_{n, \hat\alpha_n}$ is also $n^{1/2}$-consistent in a
finite interval, and its proof is provided in the online
Supplementary Material.

\section{Asymptotic properties}
Define a random map as follows
\begin{eqnarray}
&&\Psi_{\theta,n}(\phi,\alpha,\eta)= \frac{1}{n}\sum_{i=1}^n
\psi_{\theta}(Y_i,X_i,V_i,\Delta_i;\phi,\alpha,\eta),\label{lodpsin}
\end{eqnarray}
where
\begin{eqnarray*}
&& \psi_{\theta}(Y,X,V,\Delta;\phi,\alpha,\eta) \\
&& \qquad = \Delta \{Y-\dot{b}(D'(T)\theta)\}D(T)
+(1-\Delta)\left\{\int_{C-X'\alpha}^{\tau} f_{\theta,\phi}(Y |
t+X'\alpha,X)d\eta(t)\right\}^{-1}  \\
&& \qquad \qquad \quad \int_{C-X'\alpha}^{\tau} f_{\theta,\phi}(Y |
t+X'\alpha,X) \{Y-\dot{b}(D'(t+X'\alpha)\theta)\}D(t+X'\alpha)
d\eta(t) ,
\end{eqnarray*}
which is the derivative of (\ref{lodlike1}) with respect to
$\theta$. Then with $(\phi,\alpha,\eta)$ replaced by
$(\hat{\phi}_n,\hat{\alpha}_n,\hat{\eta}_{n, \hat\alpha_n})$ in
(\ref{lodpsin}), $\Psi_{\theta, n}
(\hat{\phi}_n,\hat{\alpha}_n,\hat{\eta}_{n, \hat\alpha_n}) = 0$
becomes the pseudo-likelihood estimating equation for $\theta$, and
its solution $\hat\theta_n$ is called the pseudo-likelihood
estimator.

A set of regularity conditions is introduced in the Appendix. Some
conditions are commonly assumed for the accelerated failure time
models, and other conditions are for the generalized linear models,
which are easily verifiable for linear, logistic and Poisson
regression models. We then have the following asymptotic results for
$\hat\theta_n$.

\begin{theorem} \label{Thm consist}
 (Consistency and asymptotic normality.) Denote the true value of $\theta$ by $\theta_0$. Suppose all the regularity conditions given in the Appendix hold. Then for the two-stage pseudo-likelihood estimator $\hat\theta_n$  satisfying
$\Psi_{\hat{\theta}_n,n}(\hat{\phi}_n,\hat{\alpha}_n,
\hat{\eta}_{n,\hat{\alpha}_n})=0$, we have: (i) $\hat{\theta}_n$
converges in outer probability to $\theta_0$, and (ii)
$n^{1/2}(\hat{\theta}_n-\theta_0)$ converges weakly to a mean zero
normal random variable with variance $A^{-1}BA^{-1}$, where $A$ and
$B$ are provided in the online Supplementary Material.
\end{theorem}

Because the asymptotic variance of $\hat\theta_n$ has a very
complicated expression that prohibits the direct calculation of its
estimate from observed data, we recommend using the bootstrap
variance estimator.

The proof of Theorem \ref{Thm consist} is based on the general
Z-estimation theory of \cite{nan:2012}. Define a deterministic
function
\begin{eqnarray}
\Psi_{\theta}(\phi,\alpha,\eta)=E\bigg\{\psi_{\theta}(Y,X,V,\Delta;\phi,\alpha,\eta)
\bigg\}, \label{lodpsi}
\end{eqnarray}
and denote the true values of $(\phi, \alpha, \eta)$ by $(\phi_0,
\alpha_0, \eta_0)$. We can show that
$\Psi_{\theta,n}(\hat\phi_n,\hat\alpha_n,\hat\eta_{n,\hat\alpha_n})$
converges uniformly to $\Psi_{\theta}(\phi_0,\alpha_0,\eta_0)$ as
$n\rightarrow \infty$. Then the consistency is achieved given that
$\theta_0$ is the unique solution of $\Psi_\theta(\phi_0, \alpha_0,
\eta_0)=0$. The asymptotic normality is derived by showing the
asymptotic linear representation of
$n^{1/2}(\hat{\theta}_n-\theta_0)$. The detailed proofs rely heavily
on empirical process theory and can be found in the online
Supplementary Material, where we only provide the analytic form of
the asymptotic variance for the Gehan weighted estimate of $\alpha$.
The analytic forms of the asymptotic variance for other rank based
estimates and the sieve maximum likelihood estimates can be obtained
similarly.

\section{Numerical results}
\subsection{Simulations}
We conduct simulations to investigate the finite sample performance
of the proposed method. Simulation data sets are generated from the
generalized linear model
$$
g(E(Y)) = \beta_0 + \beta_1 X_1 + \beta_2 X_2 + \gamma Z,
$$
where $\beta_0 = -1$, $\beta_1 = 0.5$, $\beta_2 = -1$, $\gamma=2$,
and $g$ is chosen to be the canonical link function for normal,
bernoulli and poisson distributions, respectively. The normal error
variance is chosen to be 1 for the linear regression model. The
three covariates are: $X_1\sim$ Bernoulli(0.5), $X_2$ is normal with
mean 1 and standard deviation 1 truncated at $\pm 3$, and $Z =
\exp{(-T)}$ is generated from the following linear model
\begin{equation*}
T=\alpha_0+\alpha_1X_1 + \alpha_2X_2+\varsigma,
\end{equation*}
where $\alpha_0=0.25$, $\alpha_1=0.25$, $\alpha_2=-0.5$,
$\varsigma\sim 0.5N(0,1/8^2)+0.5N(0.5,1/10^2)$. The limit of
detection $L$ for covariate $Z$ is chosen to yield 30\% right
censoring for $T$.

We simulate 1000 replications for each scenario and compare the
proposed method with full data analysis, complete case analysis, and
four different substitution methods. The full data analysis
represents the case of no limit of detection, which serves as a
benchmark.  We conduct simulations with two different sample sizes:
200 and 400. The four substitution methods for $Z < L$ are: (i)
replacing $Z$ by $L$, (ii) replacing $Z$ by $L/\sqrt{2}$, (iii)
replacing $Z$ by zero, and (iv) replacing $Z$ by $E(Z |  Z<L)$. For
the proposed two-stage method, we report the 90\% and 95\% coverage
proportions for which the variances are obtained from 200 bootstrap
samples. Empirical variances obtained from 1000 independent data
sets are provided only for the full data analysis and the two valid
methods for the limit of detection problem. The results are
presented in Tables \ref{tablelod:1}-\ref{tablelod:3}.

The results suggest that all the substitution methods yield biased
estimates, including substituting $Z$ by the true $E(Z |  Z<L)$. The
biases for the proposed two-stage method are minimal, which are
comparable to both the full data analysis and the complete case
analysis. Clearly, the proposed method is much more efficient than
the complete case analysis, and the bootstrap method performs well
in estimating the variance, which yields reasonable coverage rate of
the confidence intervals for all considered sample sizes.

Different censoring rates varying from 10\% to 60\% were considered
in additional simulations. Both the proposed two-stage method and
the complete case analysis yield unbiased estimates in all
simulation scenarios. The efficiency gain of the two-stage method
comparing to complete case analysis increases as the percentage of
censoring increases. All the substitution methods yield biased
estimates and the biases increase as the censoring rate increases as
well. The detailed simulation results are not provided here.

\begin{tiny}

\begin{table}[htbp!]
\caption{Simulation results for linear regression.}
\begin{footnotesize}
\begin{tabular}{lcccccc}

\\
       &          Sample size &            &  $\beta_0 = -1$  &  $\beta_1 = 0.5 $  & $\beta_2 = -1 $ &  $\gamma = 2$  \\ \\
Full data      & 200 & bias&-0.030 &0.011 &-0.006 &0.024 \\
&     & var &0.414 &0.051 &0.036 &0.374   \\
{Two-stage }     &     & {bias}&{-0.029} &{0.010} &{-0.005} &{0.022}    \\
&&{var}&{0.438} &{0.053} &{0.038} &{0.395}   \\
&&{bootstrap var}&{0.465} &{0.058} &{0.043} &{0.421} \\
&&{90\% CR (\%)}&{89.0} &{90.7 }&{91.2} &{89.0}\\
&& 95\% CR (\%) & 94.9 & 94.8 & 95.6 & 95.3 \\
Complete case      &     & bias&-0.018 &0.004 &0.000 &0.010  \\
&&var&0.531 &0.082 &0.066 &0.507\\
$L$&&bias&0.399 &-0.220 &0.228 &-0.491  \\
$L/\sqrt{2}$&&bias&0.701 &-0.136 &0.147 &-0.620 \\
Zero&&bias&1.837 &-0.417 &0.427 &-1.684 \\
$E(Z |  Z<L)$ &     & bias &0.415 &-0.133 &0.143 &-0.418 \\
\\
Full data      & 400 & bias&-0.019 &0.007 &-0.003 &0.014\\
&     & var &0.212 &0.028 &0.019 &0.192  \\
{Two-stage }     &     & {bias}&{-0.019} &{0.008} &{-0.003} &{0.015}  \\
&&{var}&{0.225} &{0.029} &{0.020} &{0.204}  \\
&&{bootstrap var}&{0.226 }&{0.028} &{0.021} &{0.205}  \\
&&{90\% CR (\%)}&{89.4}&{89.2}&{90.4}&{89.9} \\
&& 95\% CR (\%) & 95.0 & 93.8 & 95.0 & 95.0 \\
Complete case      &     & bias&-0.019 &-0.001 &0.004 &0.008  \\
&&var&0.273 &0.043 &0.033 &0.255   \\
$L$&&bias&0.404 &-0.221 &0.230 &-0.495 \\
$L/\sqrt{2}$&&bias&0.724 &-0.144 &0.155 &-0.642   \\
Zero&&bias&1.850 &-0.426 &0.436 &-1.700  \\
$E(Z |  Z<L)$ &     & bias &0.433 &-0.138 &0.148 &-0.434\\
\end{tabular}
\label{tablelod:1}
\end{footnotesize}
\end{table}

\begin{table}[htbp!]
\caption{Simulation results for logistic regression.}
\begin{footnotesize}
\begin{tabular}{lcccccc}
\\
       &          Sample size &            &  $\beta_0 = -1$  &  $\beta_1 = 0.5 $  & $\beta_2 = -1 $ &  $\gamma = 2$  \\
\\
Full data      & 200 & bias&-0.030 &0.013 &-0.033 &0.060 \\
&     & var &2.157 &0.268 &0.216 &2.033  \\
{Two-stage  }    &     & {bias}&{-0.041} &{0.016} &{-0.037} &{0.071}   \\
&&{var}&{2.313} &{0.278} &{0.230} &{2.191}   \\
&&{bootstrap var}&{2.424} &{0.299} &{0.235} &{2.260}  \\
&&{90\% CR (\%)}&{91.4} &{91.7
}&{90.7} &{91.0}\\
&& 95\% CR (\%) & 96.2 & 96.3 & 95.9 & 96.2 \\
Complete case      &     & bias&-0.076 &0.021 &-0.045 &0.106\\
&&var&2.822 &0.413 &0.381 &2.842 \\
$L$&&bias&0.309 &-0.185 &0.171 &-0.368\\
$L/\sqrt{2}$&&bias&0.690 &-0.122 &0.110 &-0.570 \\
Zero&&bias&1.880 &-0.453 &0.441 &-1.716 \\
$E(Z |  Z<L)$ &     & bias &0.350 &-0.100 &0.087 &-0.313  \\
\\
Full data      & 400 & bias&-0.033 &0.007 &-0.016 &0.041\\
&     & var &0.930 &0.123 &0.096 &0.881 \\
{Two-stage}      &     & {bias}&{-0.043} &{0.011} &{-0.020} &{0.052}   \\
&&{var}&{1.013} &{0.129} &{0.104 }&{0.964} \\
&&{bootstrap var}&{1.101} &{0.138} &{0.107} &{1.022}  \\
&&{90\% CR
(\%)}&{90.8}&{91.2}&{90.5}&{90.6}\\
&& 95\% CR (\%) & 95.8 & 96.3 & 95.8 & 95.2 \\
Complete case      &     & bias&-0.037 &0.005 &-0.018 &0.048  \\
&&var&1.169 &0.190 &0.159 &1.160  \\
$L$&&bias&0.319 &-0.193 &0.190 &-0.398 \\
$L/\sqrt{2}$&&bias&0.651 &-0.119 &0.117 &-0.553    \\
Zero&&bias&1.841 &-0.442 &0.440 &-1.691  \\
$E(Z |  Z<L)$ &     & bias &0.332 &-0.103 &0.101 &-0.317 \\
\\
\end{tabular}
\label{tablelod:2}
\end{footnotesize}
\end{table}

\begin{table}[htbp!]
\caption{Simulation results for Poisson regression.}
\begin{footnotesize}
\begin{tabular}{lcccccc}
\\
       &          Sample size &            &  $\beta_0 = -1$  &  $\beta_1 = 0.5 $  & $\beta_2 = -1 $ &  $\gamma = 2$  \\
\\
Full data      & 200 & bias&0.022 &-0.009 &0.008 &-0.026  \\
&     & var &0.225 &0.024 &0.018 &0.198    \\
{Two-stage }     &     & {bias}&{0.034 }&{-0.011 }&{0.011 }&{-0.037 }  \\
&&{var}&{0.250 }&{0.026} &{0.020} &{0.221}  \\
&&{bootstrap var}&{0.249} &{0.027} &{0.020} &{0.218}   \\
&&{90\% CR(\%)}&{90.9} &{89.9
}&{90.0} &{90.6}  \\
&& 95\% CR (\%) & 94.5 & 94.8 & 94.7 & 94.8 \\
Complete case      &     & bias&0.025 &-0.011 &0.010 &-0.031\\
&&var&0.351 &0.053 &0.041 &0.325  \\
$L$&&bias&0.589 &-0.288 &0.286 &-0.660 \\
$L/\sqrt{2}$&&bias&0.885 &-0.200 &0.210 &-0.801  \\
Zero&&bias&1.867 &-0.380 &0.396 &-1.691 \\
$E(Z |  Z<L)$ &     & bias &0.637 &-0.213 &0.217 &-0.628 \\
\\
Full data      & 400 & bias&0.018 &-0.003 &0.005 &-0.020 \\
&     & var &0.105 &0.012 &0.008 &0.092 \\
{Two-stage }     &     & {bias}&{0.019} &{-0.003} &{0.005 }&{-0.021}  \\
&&{var}&{0.119} &{0.013} &{0.009} &{0.104}  \\
&&{bootstrap var}&{0.121} &{0.013} &{0.010} &{0.105}  \\
&&{90\% CR
(\%)}&{90.1}&{90.7}&{90.5}&{90.7}\\
&& 95\% CR (\%) & 95.2 & 95.3 & 94.8 & 95.0 \\
Complete case      &     & bias&0.016 &-0.004 &0.007 &-0.022  \\
&&var&0.175 &0.027 &0.022 &0.163   \\
$L$&&bias&0.578 &-0.281 &0.283 &-0.649  \\
$L/\sqrt{2}$&&bias&0.886 &-0.196 &0.208 &-0.800    \\
Zero&&bias&1.870 &-0.373 &0.391 &-1.689  \\
$E(Z |  Z<L)$ &     & bias &0.633 &-0.208 &0.215 &-0.623  \\
\\
\end{tabular}
\label{tablelod:3}
\end{footnotesize}
\end{table}

\end{tiny}
The consistency of the estimates from the two-stage method depends
on correctly specifying the semiparametric accelerated failure time
model (\ref{lodAFT}). Although the assumption for model
(\ref{lodAFT}) is much less restrictive than any parametric model,
model checking should be done before applying the proposed two-stage
method. Though less severe than ad hoc substitution methods,
misspecification of the accelerated failure time model also yields
biased results, and the bias increases as the severity of
misspecification of (\ref{lodAFT}) grows. Again, the detailed
simulation results are not provided here.

\subsection{The National Health and
Nutrition Examination Survey}

We consider the National Health and Nutrition Examination Survey
2003-2004 as an illustrative example for the regression with a
covariate subject to limit of detection. In particular, we focus on
the effect of left-censored urine arsenobetaine on the prevalence of
type 2 diabetes, see \cite{Navas-Acien:2008}

National Health and Nutrition Examination Survey, conducted by the
US National Center for Health Statistics, used a complex multistage
sampling design to obtain a representative sample of the civilian
noninstitutionalized individuals within the US population. The data
set contains a subsample of 1542 study participants with arsenic
measurements of the National Health and Nutrition Examination Survey
2003-2004. For each participant in this subsample, total urine
arsenic and arsenic species including arsenobetaine were collected
for arsenic analysis. The limit of detection for total urine arsenic
and urine arsenobetaine were 0.6 and 0.4 $\mu$g/l, respectively. The
percentage of study participants with levels below the limit of
detection were 1.3\% for total urine arsenic and 27.8\% for urine
arsenobetaine \citep{Caldwell2009}.  \cite{Navas-Acien:2008} found
that total urine arsenic was associated with increased prevalence of
type 2 diabetes, hence it was adjusted in the analysis, and 19
participants with total urine arsenic below limit of detection were
dropped from the study. The urine creatinine level that was used to
account for urine dilution in spot urine samples was fully observed
and was adjusted in the analysis as well. We further excluded 23
participants with missing values in other variables of interest.

\begin{figure}
\caption{Goodness of fit for the AFT model }
  \begin{center}
    \resizebox{120mm}{!}{\includegraphics{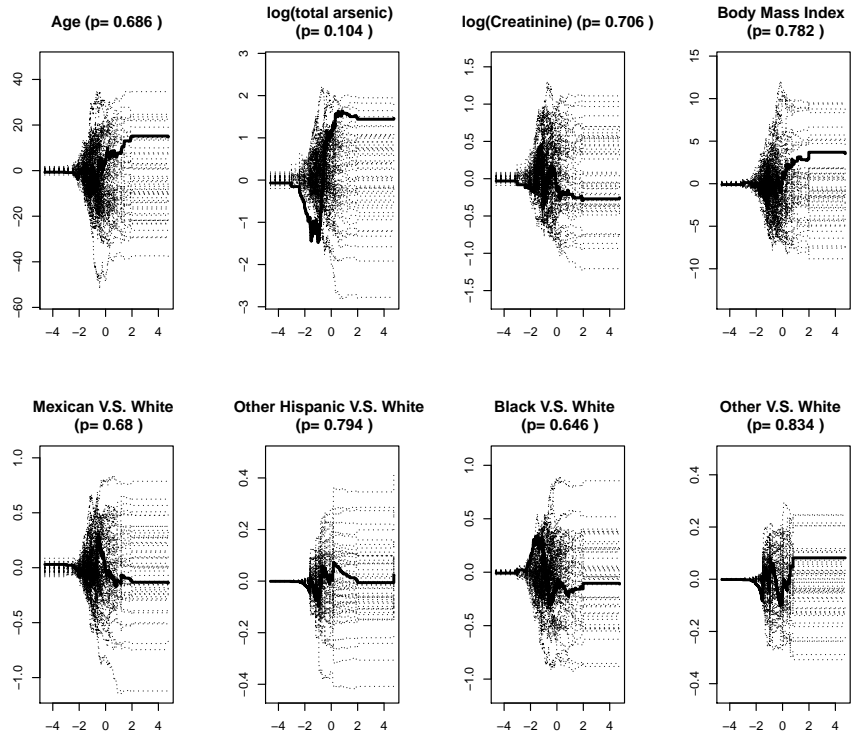}} \hspace{15mm}
    \label{GOF}
  \end{center}
\end{figure}

For illustration purpose, we focus on the male participants, where
24.1\% of 730 subjects had urine arsenobetaine below limit of
detection. Age, race/ethnicity, body mass index, the logarithm of
total urine arsenic and the logarithm of urine creatinine level are
used as covariates to fit the accelerated failure time model for
$-\log(\mbox{arsenobetaine})$. All of them are significant with
p-values $<0.0001$ except one dummy variable for race. Dotted lines
in Figure \ref{GOF} are the plots of 50 realizations from the
distributions of the score processes. The observed score processes
are presented with solid lines which randomly fluctuate around zero.
From Figure \ref{GOF} we see that the considered accelerated failure
time model for urine arsenobetaine fits the data reasonably well,
with respective goodness-of-fit empirical p-values of 0.686, 0.104,
0.706, 0.782, 0.68, 0.794, 0.646 and 0.834 for age, log total urine
arsenic, log urine creatinine, body mass index and race that
contains five ethnic groups based on 500 simulated martingale
residual score processes. The response of interest is the status of
type 2 diabetes. Thus, a logistic regression is considered with age,
body mass index, log total urine arsenic, log urine creatinine and
log urine arsenobetaine as the covariates of interest, whereas race
is not significantly associated. Table \ref{realtable} shows the
regression coefficient estimates, where we see that the proposed
two-stage method yields similar point estimates with smaller
variances/p-values comparing to the complete case analysis,
indicating the efficiency gain of the proposed method.
Substitution methods with values below limit of detection replaced
by $L$ or $L/\sqrt{2}$ yield quite different point estimates
comparing to the complete case analysis and the two-stage method,
thus put the significant results of $\log(\mbox{arsenobetaine})$ in
question. The effect of $\log(\mbox{arsenobetaine})$ is clearly
biased when the urine arsenobetaine values below limit of detection
are replaced by zero.

\begin{table}[htbp]
\caption{Regression analysis results for the prevalence of type 2
diabetes with covariate urinary arsenobetaine subject to limit of
detection: the National Health and Nutrition Examination Survey
2003-2004.}
\begin{tiny}
\begin{tabular}{lcccccc}
\\
       &              &        Age  & log(Arsenic) &log(Creatinine) &BMI & log(Arsenobetaine)  \\
{Two-stage  }    &    {estimate} &{0.05} &{0.50}& -0.76& 0.10&-0.22   \\
&{bootstrap sd}&{0.007} &{0.25}&0.27&0.02&0.12   \\
& p-value & $<0.0010$ & 0.04&0.004& $<0.0001$&0.07\\
Complete case     &      estimate &0.06 &0.41&-0.81&0.13&-0.20  \\
&sd &0.01 &0.32&0.33&0.03&0.20   \\
& p-value & $<0.0010$ & 0.20& 0.01&$<0.0001$ &0.31\\
$L$      &      estimate &0.05 &0.57 & -0.81& 0.10& -0.32 \\
&sd &0.008 &0.25 & 0.26&0.02&0.15   \\
& p-value & $<0.0001$ & 0.02 & 0.002& $<0.0001$& 0.03\\
$L/\sqrt{2}$      &      estimate &0.05 &0.56 & -0.80 & 0.10 & -0.30 \\
&sd &0.008 &0.24 & 0.26 & 0.02 & 0.14   \\
& p-value & $<0.0001$ & 0.02 & 0.002 & $<0.0001$ & 0.02\\
Zero      &      estimate &0.05 &0.25 & -0.61 & 0.10 & -0.03 \\
&sd &0.008 &0.16 & 0.24 & 0.02 & 0.02   \\
& p-value & $<0.0001$ & 0.12 & 0.009 & $<0.0001$ &0.11\\
\end{tabular}
\label{realtable}
\end{tiny}
\end{table}

\section{Disscussion}


The estimates from the proposed method are consistent and
asymptotically normal under much less restrictive assumptions than
parametric approaches. Another advantage of the two-stage method
over the parametric method is that the model checking tools are
available. All the substitution methods could yield large bias
including replacing the values below the limit of detection by the
true $E(Z|Z<L)$. The consistency and efficiency gain of the proposed
two-stage method rely on the correctly specified accelerated failure
time model. We suggest to fit the accelerated failure time model for
the covariate subject to limit of detection before applying the
two-stage method. If the data fits the model reasonably well and
there are at least some fully observed covariates significantly
associated with the covariate subject to limit of detection, then
the proposed method is recommended. Otherwise, we suggest to use the
complete case analysis.

The amount of efficiency gain of the proposed two-stage method
depends on how far we can estimate $F(t | X)$ reasonably well beyond
the limit of detection. We assume some finite value $\tau$ for
residuals in this article. In practice, the upper limit of the
integral in the pseudo-likelihood function can go as far as the
largest observed residual in the fitted accelerated failure time
model, which is $\max_i(T_i - X_i'\hat\alpha_n)$;
 and
theoretically, this upper limit is $\infty$ when the support of
$X'\alpha_0$ is unbounded. In the latter case, it can be shown that
$\hat{F}_n$ converges to $F$ on the entire real line with a
polynomial rate at $n^{-1/8}$, see \cite{lai:1991} and \cite{Ding
and Nan:2014}, and we may still obtain consistent estimates for the
parameters of interest. The asymptotic normality, however, will be
difficult to show.

We only consider the case with one covariate subject to limit of
detection in this article for simplicity. Regression with multiple
covariates subject to limits of detection may occur in practice.
Parametric models have been considered for such problems
\citep{may:2011, D'Angelo and Weissfeld:2008}. To achieve robust
results, the proposed semiparametric approach can be generalized to
tackle the problem with multiple covariates subject to limits of
detection. The critical step is to provide an valid nonparametric
estimate for the multivariate survival function, for which available
methods include \cite{dabrowska:1988}, \cite{prentice:1992},
\cite{varderlaan:1996}, and \cite{prentice:2004}. The constant limit
of detection assumption considered in this article, though commonly
seen in practice, is also for notational simplicity, and can be
relaxed to cases with random limit of detection.

Limit of detection issue is a special missing data problem. Multiple
imputation \citep{little rubin:2002} may be considered as an
alternative method if the tail distribution of the covariate subject
to limit of detection conditional on all other variables, including
the response variable, can be estimated reasonably well. This
research is ongoing and will be presented elsewhere.

When the original lab measurements are available, which may or may
not be below limit of detection,  \cite{Murphy2010 } and
\cite{Louis2012 } directed used the machine-read lab values in their
analysis to avoid the potential biases caused by substitution. More
appropriate analysis should be treating these lab data as error
prone measurements \citep{Guo:2010}, hence methods dealing with
measurement errors would apply.

The proposed two-stage method is ready to be generalized to other
regression models that have a likelihood function to work with, for
example, the Cox regression model and mixed effects model. Extra
care will be needed due to special features of these models. For
example, handling the nonparametric baseline function in the Cox
model in the context of limit of detection can be delicate. These
are interesting topics worth further investigation.

\section*{Supplementary material}

The online Supplementary Material contains the detailed proof of
Theorem 1.

\section*{Acknowledgement}

We thank Professor Sung Kyun Park for the National Health and
Nutrition Examination Survey arsenic exposure data. We are grateful
to the editor, associate editor and a referee for helpful and
constructive comments. The research was partially supported by
several grants from the U.S. National Science Foundation and the
National Institute of Health.

\section*{Appendix: regularity conditions}

Denote the sample space of response variable $Y$ by $\mathcal{Y}$,
the sample space of covariate $X$ by $\mathcal{X}$, the parameter
space of $\theta$ by $\Theta$, the parameter space of $\alpha$ by
$\mathcal{A}$, and the parameter space of $\eta$ by $\mathcal{H}$.
In addition to the assumptions of bounded support for $(X,Z)$ and
compact parameter spaces $\Theta$ and ${\cal A}$, we provide a set
of  regularity conditions for Theorem \ref{Thm consist} in the
following.

\begin{condition}\label{condition1}
$\Psi_{\theta}(\phi_0,\alpha_0,\eta_{0,\alpha_0})$ has a unique root
$\theta_0$.
\end{condition}

\begin{condition}\label{condition2}
For any constant $U<\infty$, $\sup_{t\in[C,U]}|h(t)|\leq E_0<\infty$
$\sup_{t\in[C,U]}|\dot{h}(t)|\leq E_1<\infty$, and
$\sup_{t\in[C,U]}|\ddot{h}(t)|\leq E_2<\infty$, where $\dot{h}$ and
$\ddot{h}$ are the first and second derivatives of $h$ respectively,
and $E_0$, $E_1$ and $E_2$ are constants.
\end{condition}

\begin{condition}\label{condition3}
 Error $\varsigma$ has bounded density
$f=\dot\eta_{0,\alpha_0}$ with bounded derivative $\dot{f}$, in
other words, $f\leq E_3<\infty$, $|\dot{f}|\leq E_4<\infty$ for
constants $E_3$ and $E_4$, and
\begin{equation*}
\int_{-\infty}^{\infty}(\dot{f}(t)/f(t))^2f(t)dt<\infty.
\end{equation*}
\end{condition}

\begin{condition}\label{condition4}
 There is a constant $\tau<\infty$ such that
$pr(V-X'\alpha\geq \tau|X=x)>\xi>0$ for all $x\in\mathcal{X}$ and
$\alpha\in \mathcal{A}$.
\end{condition}

\begin{condition}\label{condition5}
$a(\phi)$ is a monotone function satisfying $|1/a(\phi)|\leq
l<\infty$ for a constant $l$ with bounded derivatives
$\dot{a}(\cdot)$ and $\ddot{a}(\cdot)$.
\end{condition}

\begin{condition}\label{condition6}
$\dot{b}(\cdot)$ is a bounded monotone function.
\end{condition}

\begin{condition}\label{condition7}
$\ddot{b}(\cdot)$ is a bounded Lipschitz function.
\end{condition}

\begin{condition}\label{condition8}
There exist constants $C_i$, $i=1,\dots,5$,  such that for any
constant $U<\infty$,
\begin{eqnarray*}
&&\sup_{y\in \mathcal{Y}, \theta\in\Theta,|1/a(\phi)|\leq
l,x\in\mathcal{X},t\in [C,U]}\left|f_{\theta,\phi}(y |
t,x)\{y-\dot{b}(D'(t)\theta)\}\right|\leq
C_1<\infty,\\
&&\sup_{y\in \mathcal{Y}, \theta\in\Theta,|1/a(\phi)|\leq
l,x\in\mathcal{X},t\in [C,U]}\left|\frac{\partial f_{\theta,\phi}(y
|  t,x)}{\partial
\phi}\{y-\dot{b}(D'(t)\theta)\}\right|\leq C_2<\infty,\\
&&\sup_{y\in \mathcal{Y}, \theta\in\Theta,|1/a(\phi)|\leq
l,x\in\mathcal{X},t\in [C,U]}\left|\frac{\partial
\left[f_{\theta,\phi}(y |  t,x)\{y-\dot{b}(D'(t)\theta)\}\right]}{\partial t}\right|\leq C_3<\infty,\\
&&\sup_{y\in \mathcal{Y}, \theta\in\Theta,|1/a(\phi)|\leq
l,x\in\mathcal{X},t\in [C,U]}\left|\frac{\partial f_{\theta,\phi}(y
|  t,x)}{\partial
\phi}\right|\leq C_4<\infty,\\
&&\sup_{y\in \mathcal{Y}, \theta\in\Theta,|1/a(\phi)|\leq
l,x\in\mathcal{X},t\in [C,U]}\left|\frac{\partial
\left[f_{\theta,\phi}(y |
t,x)\{y-\dot{b}(D'(t)\theta)\}\right]}{\partial \theta}\right|\leq
C_5<\infty.
\end{eqnarray*}
\end{condition}

\begin{condition}\label{condition9}
There exist constants $\delta_1>0$ and $\delta_2>0$, such
that$\int_{C-X'\alpha}^{\tau}f_{\theta,\phi}(Y |
t+X'\alpha,X)d\eta(t)\geq\delta_1$ with probability 1 for any
$\theta\in\Theta$ and
$|\phi-\phi_0|+|\alpha-\alpha_0|+\|\eta-\eta_0\|<\delta_2$.
\end{condition}

{\sc Remark}: Condition \ref{condition1} is for the consistency,
which may be  unnecessarily strong for the proposed two-stage
method. Direct calculation yields
\begin{eqnarray*}
\dot{\Psi}_{\theta_0} &=&\frac{\partial
\Psi_{\theta}(\phi_0,\alpha_0,\eta_{0})}{\partial
\theta}\bigg|_{\theta=\theta_0}\\
&=& E\bigg\{-\Delta \ddot{b}\{D'(T)\theta_0\}D(T)^{\otimes
2}-(1-\Delta)\bigg(\int_{C-X'\alpha_0}^{\tau} f_{\theta_0,\phi_0}(Y|
t+X'\alpha_0,X)d\eta_{0}(t)\bigg)^{-
2}\\
&& \bigg(\int_{C-X'\alpha_0}^{\tau}  f_{\theta_0,\phi_0}(Y|
t+X'\alpha_0,X) [Y-\dot{b}\{D'(t+X'\alpha_0)\theta_0\}]
D(t+X'\alpha_0)d\eta_{0}(t)\bigg)^{\otimes 2} \bigg\},
\end{eqnarray*}
which is negative definite. Thus $\dot\Psi_\theta$, a continuous
matrix with $\theta$, is also negative definite in a neighborhood of
$\theta_0$, which guarantees that  $\theta_0$ is the unique solution
of $\Psi_{\theta}(\phi_0,\alpha_0,\eta_{0})=0$   in a neighborhood
of $\theta_0$.  The initial value we use in the Newton-Raphson
algorithm for solving
$\Psi_{\theta,n}(\hat{\phi}_n,\hat{\alpha}_n,\hat{\eta}_{n,\hat{\alpha}_n})=0$
is obtained from the complete case analysis, which is consistent,
thus the solution of the proposed two-stage method should also be
consistent.

Condition \ref{condition2} holds for many commonly used
transformations, for example, $h(t)=\exp(-t)$ and polynomial
functions. Condition \ref{condition3} and \ref{condition4} are usual
assumptions for accelerated failure time models \citep{Tsiatis:1990,
nan:2009}. Conditions \ref{condition5}-\ref{condition8}
automatically hold for common generalized linear models, for
example, linear, logistic or poisson regression.

 Condition
\ref{condition9} is mainly for technical convenience. One way to
obtain Condition \ref{condition9} might be to truncate response
variable $Y$ such that $|Y|\leq M<\infty$ for a large constant $M$
and to further truncate the residual in the accelerated failure time
model with some constant $\tau'<\tau$. In our simulations, however,
we do not implement such truncations but still obtain satisfactory
results.

\end{document}


\title{Semiparametric Approach for
Regression with Covariate Subject to Limit of Detection\\
(Supplementary Data)}

\author{Shengchun Kong and Bin Nan}

\maketitle
\section{General Z-estimation theory}

The proof of Theorem 1 in the main text is based on the general
Z-estimation theory of \cite{nan:2012}, which is provided in the
following Lemmas \ref{Lemma consist} and \ref{Lemma an} for our
problem setting. Detailed discussion and proofs of these two lemmas
can be found in \cite{nan:2012}. Let $|\cdot|$ be the Euclidian norm
and $\|\eta-\eta_{0}\| =\sup_{t}|\eta(t)-\eta_{0}(t)|$. Define
$\rho\{(\phi,\alpha,\eta),(\phi_0,\alpha_0,\eta_{0})\}=|\phi-\phi_0|+|\alpha-\alpha_0|+\|\eta-\eta_{0}\|$.
We use $P^*$ to denote outer probability, which is defined as
$P^*(A)=\inf\{pr(B):B\supset A,B \in \mathcal{B}\}$ for any subset
$A$ of $\Omega$ in a probability space $(\Omega,\mathcal{B},P)$.

\begin{lemma} \label{Lemma consist}
(Consistency.) Suppose $\theta_0$ is the unique solution to
$\Psi_{\theta}(\phi_0,\alpha_0,\eta_{0})=0$ in the parameter space
$\Theta$ and ($\hat{\phi}_n, \hat{\alpha}_n,
\hat{\eta}_{n,\hat{\alpha}_n}$) are estimators of ($\phi_0,
\alpha_0, \eta_{0}$) such that $\rho\{(\hat{\phi}_n,
\hat{\alpha}_n,\hat{\eta}_{n,\hat{\alpha}_n}),(\phi_0,
\alpha_0,\eta_{0})\}=o_{p^*}(1)$. If
\begin{equation}
\sup_{\theta\in\Theta,\rho\{(\phi,\alpha,\eta),(\phi_0,\alpha_0,\eta_{0})\}\leq
\delta_n}\frac{|\Psi_{n,\theta}(\phi,\alpha,\eta)-\Psi_{\theta}(\phi_0,\alpha_0,\eta_{0})|}
{1+|\Psi_{n,\theta}(\phi,\alpha,\eta)|+|\Psi_{\theta}(\phi_0,\alpha_0,\eta_{0})|}=o_{p^*}(1)\label{lemma1}
\end{equation}
for every sequence $\{\delta_n\downarrow 0\}$, then $\hat{\theta}_n$
satisfying
$\Psi_{n,\hat{\theta}_n}(\hat{\phi}_n,\hat{\alpha}_n,\hat{\eta}_{n,\hat{\alpha}_n})=o_{p^*}(1)$
converges in outer probability to $\theta_0$.
\end{lemma}

\begin{lemma} \label{Lemma an} (Rate of convergence and asymptotic representation.)
Suppose that $\hat{\theta}_n$ satisfying
$\Psi_{n,\hat{\theta}_n}(\hat{\phi}_n,\hat{\alpha}_n,\hat{\eta}_{n,\hat{\alpha}_n})=o_{p^*}(n^{-1/2})$
is a consistent estimator of $\theta_0$ that is a solution to
$\Psi_{\theta}(\phi_0,\alpha_0,\eta_{0})=0$ in $\Theta$, and that
($\hat{\phi}_n, \hat{\alpha}_n,\hat{\eta}_{n,\hat{\alpha}_n}$) is an
estimator of ($\phi_0, \alpha_0, \eta_{0}$) satisfying
$\rho\{(\hat{\phi}_n,\hat{\alpha}_n,\hat{\eta}_{n,\hat{\alpha}_n}),$
$(\phi_0,\alpha_0,\eta_{0})\}=O_{p^*}(n^{-1/2})$. Suppose the
following four conditions are satisfied:

\medskip

(i) (Stochastic equicontinuity.)
\begin{equation*}
\frac{|n^{1/2}(\Psi_{n,\hat{\theta}_n}-\Psi_{\hat{\theta}_n})(\hat{\phi}_n,\hat{\alpha}_n,\hat{\eta}_{n,\hat{\alpha}_n})
-n^{1/2}(\Psi_{n,\theta_0}-\Psi_{\theta_0})(\phi_0,\alpha_0,\eta_{0})|}
{1+n^{1/2}|\Psi_{n,\hat{\theta}_n}(\hat{\phi}_n,\hat{\alpha}_n,\hat{\eta}_{n,\hat{\alpha}_n})|
+n^{1/2}|\Psi_{\hat{\theta}_n}(\hat{\phi}_n,\hat{\alpha}_n,\hat{\eta}_{n,\hat{\alpha}_n})|}=o_{p^*}(1).
\end{equation*}

\medskip

(ii)
$n^{1/2}\Psi_{n,\theta_0}(\phi_0,\alpha_0,\eta_{0})=O_{p^*}(1)$.

\medskip

(iii) (Smoothness.) There exist continuous matrices
$\dot{\Psi}_{1,\theta_0}(\phi_0,\alpha_0,\eta_{0})$,
$\dot{\Psi}_{2,\theta_0}(\phi_0,\alpha_0,\eta_{0})$,
$\dot{\Psi}_{3,\theta_0}(\phi_0,\alpha_0,\eta_{0})$, and a
continuous linear functional
$\dot{\Psi}_{4,\theta_0}(\phi_0,\alpha_0,\eta_{0})$  such that
\begin{eqnarray}
&&|\Psi_{\hat{\theta}_n}(\hat{\phi}_n,\hat{\alpha}_n,\hat{\eta}_{n,\hat{\alpha}_n})
-\Psi_{\theta_0}(\phi_0,\alpha_0,\eta_{0})\nonumber\\
&& \qquad -\dot{\Psi}_{1,\theta_0}(\phi_0,\alpha_0,\eta_{0})
(\hat{\theta}_n-\theta_0)-\dot{\Psi}_{2,\theta_0}(\phi_0,\alpha_0,\eta_{0})(\hat{\phi}_n-\phi_0)\nonumber\\
&& \qquad -\dot{\Psi}_{3,\theta_0}(\phi_0,\alpha_0,\eta_{0})
(\hat{\alpha}_n-\alpha_0)
-\dot{\Psi}_{4,\theta_0}(\alpha_0,\eta_{0})(\hat{\eta}_{n,\hat{\alpha}_n}-\eta_{0})|
\nonumber\\
&&
=o(|\hat\theta_n-\theta_0|)+o[\rho\{(\hat{\phi}_n,\hat{\alpha}_n,\hat{\eta}_{n,\hat{\alpha}_n}),(\phi_0,\alpha_0,\eta_{0})\}]\label{lemma21}.
\end{eqnarray}
Here the subscripts 1, 2, 3, and 4 correspond to $\theta$, $\phi$,
$\alpha$, and $\eta$ in $\Psi_\theta(\phi,\alpha,\eta)$,
respectively, and
 we assume that
the matrix $\dot{\Psi}_{1,\theta_0}(\phi_0,\alpha_0,\eta_{0})$ is
nonsingular.

\medskip

(iv) $n^{1/2}\dot{\Psi}_{2,\theta_0}(\phi_0,\alpha_0,\eta_{0})
(\hat{\phi}_n-\phi_0)=O_{p^*}(1)$,
$n^{1/2}\dot{\Psi}_{3,\theta_0}(\phi_0,\alpha_0,\eta_{0})(\hat{\alpha}_n-\alpha_0)=O_{p^*}(1)$,
and $n^{1/2}\dot{\Psi}_{4,\theta_0}(\phi_0,\alpha_0,\eta_{0})
(\hat{\eta}_{n,\hat{\alpha}_n}-\eta_{0})=O_{p^*}(1)$.

\medskip

Then $\hat{\theta}_n$ is $n^{1/2}$ -consistent and further we have
\begin{eqnarray}
n^{1/2}(\hat{\theta}_n-\theta_0)&=&\{-\dot{\Psi}_{1,\theta_0}(\phi_0,\alpha_0,\eta_{0})\}^{-1}n^{1/2}
\{(\Psi_{n,\theta_0}-\Psi_{\theta_0})(\phi_0,\alpha_0,\eta_{0}) \nonumber \\
&& \qquad +\dot{\Psi}_{2,\theta_0}(\phi_0,\alpha_0,\eta_{0})
(\hat{\phi}_n-\phi_0)+\dot{\Psi}_{3,\theta_0}(\phi_0,\alpha_0,\eta_{0})
(\hat{\alpha}_n-\alpha_0)\nonumber\\
&& \qquad +\dot{\Psi}_{4,\theta_0}(\phi_0,\alpha_0,\eta_{0})
(\hat{\eta}_{n,\hat{\alpha}_n}-\eta_{0})\}+o_{p^*}(1).\label{lemma23}
\end{eqnarray}
\end{lemma}

\medskip

\section{Technical lemmas}

Now we provide technical preparations for the proof of Theorem 1,
some of which are from Ying Ding's 2010 University of Michigan Ph.D.
thesis. We adopt the empirical process notation of \cite{van and
wellner:1996}.

Let $\epsilon_\alpha=V-X'\alpha$ and $\epsilon_0=V-X'\alpha_0$.
Define
\begin{eqnarray*}
h^{(0)}(\alpha,s) &=& P\{1(\epsilon_{\alpha}\leq
s,\Delta=1)\}, \\
h^{(1)}(\alpha,s) &=& P\{1(\epsilon_{\alpha}\geq s)\}, \\
h^{(2)}(\alpha,s) &=& P\{1(\epsilon_{\alpha}\geq s)X\},
\end{eqnarray*}
and
$$
H^{(1)}_n(\alpha,s) = {\mathbb P}_n\{1(\epsilon_{\alpha}\geq s)\}.
$$
The Kaplan-Meier estimator of the distribution function of $T-\alpha
X$ is given by
\begin{equation*}
\hat{\eta}_{n,\alpha}(t)=1-\prod_{i: V_i-X_i'\alpha\leq
t}\left\{1-\frac{\Delta_i/n}{H^{(1)}_n(\alpha,V_i-X_i'\alpha)}\right\}.
\end{equation*}
Define
\begin{equation*}
F(\alpha,t)=1-\exp\bigg\{-\int_{u\leq t}\frac{d
h^{(0)}(\alpha,u)}{h^{(1)}(\alpha,u)}\bigg\},
\end{equation*}
and denote $ \dot{F}_{\alpha}(\alpha,t)={\partial
F(\alpha,t)}/{\partial \alpha}. $ For function $c$ in the
exponential family, denote $ \dot{c}_{\phi}(Y,\phi_0)={\partial
c(Y,\phi)}/{\partial \phi}|_{\phi=\phi_0}. $

Let
$\Phi\{\alpha,h^{(1)},h^{(2)}\}=P\left[\left\{h^{(1)}(\alpha,\epsilon_{\alpha})X-h^{(2)}(\alpha,\epsilon_{\alpha})\right\}\Delta\right]$,
which corresponds to the limiting Gehan weighted estimating
function, and define

\begin{eqnarray}
&&m_1(\alpha_0,s;t)=-P\bigg\{\frac{\Delta 1(s\geq
\epsilon_{0})1(t\geq
\epsilon_{0})}{h^{(1)}(\alpha_0,\epsilon_{0})^2}\bigg\},~~m_2(\alpha_0,s;t,\Delta)=\frac{\Delta
1(t\geq s)}{h^{(1)}(\alpha_0,s)},\label{appb3} \\
&&m_3(\alpha_0,\epsilon_0;\Delta,X) \nonumber\\
&&\hspace{5mm}=\bigg[-\dot{\Phi}_{\alpha}\{\alpha_0,h^{(1)}(\alpha_0,\cdot),h^{(2)}(\alpha_0,\cdot)\}
\bigg]^{-1}\bigg[\big\{h^{(1)}(\alpha_0,\cdot)X-h^{(2)}(\alpha_0,\cdot)\big\}\Delta\label{appb4}
\\
&&\hspace{10mm}-\int\{1(\epsilon_0\geq
t)X\}dP_{\epsilon_0,\Delta}(t,1)+\int\{1(\epsilon_0\geq
t)\}xdP_{\epsilon_0,\Delta,X}(t,1,x)\bigg].\nonumber
\end{eqnarray}

\medskip

\begin{lemma}\label{rootnconsis}
Suppose Conditions 3-4 hold, and let $\hat\alpha_n$ be the Gehan
weighted estimator for $\alpha_0$, we have
\begin{equation*}
\sup_{t\in[C-E_5,\tau]}|\hat{\eta}_{n,\hat{\alpha}_n}(t)-\eta_{0}(t)|=O_{p^*}(n^{-1/2}),
\end{equation*}
where $C$ is transformed $L$ and $E_5=\sup_{\alpha\in{\cal
A},x\in\mathcal{X}}|x'\alpha| < \infty$.
\end{lemma}

\begin{proof}
From the proof of Ying Ding's Theorem 2.2.3 in her 2010 University
of Michigan Ph.D. thesis, for $t$ in a bounded interval, we have for
$t\in[C-E_5,\tau]$,
\begin{eqnarray}
\sup_{t} n^{1/2}\{\hat{\eta}_{n,\hat{\alpha}_n}(t)-\eta_{0}(t)\} &=&
\sup_{t}\mathbb{G}_n[\{1-\eta_{0}(t)\}\{m_1(\alpha_0,\epsilon_0;t)+m_2(\alpha_0,\epsilon_0;t,\Delta)\}
\nonumber\\
&& \quad
+\dot{F}_{\alpha}(\alpha_0,t)m_3(\alpha_0,\epsilon_0;X,\Delta)]+o_p(1),
\label{eta}
\end{eqnarray}
where $m_1(\alpha_0,s;t)$, $m_2(\alpha_0,\epsilon_0;t,\Delta)$,
$m_3(\alpha_0,\epsilon_0;X,\Delta)$ are defined in (\ref{appb3}) and
(\ref{appb4}).

We first calculate the bracket numbers for
$\mathcal{F}_1=\{m_1(\alpha_0,\epsilon_0;t), t\in[C-E_5,\tau]\}$ and
$\mathcal{F}_2=\{m_2(\alpha_0,\epsilon_0;t,\Delta),
t\in[C-E_5,\tau]\}$. For any nontrivial $\varepsilon$ satisfying
$1>\varepsilon>0$, let $t_i$ be the i-th $\lceil 1/\varepsilon
\rceil$ quantile of $\varsigma_{0}=T-X'\alpha_0$, i.e.
\begin{equation*}
pr(\varsigma_{0}\leq t_i)=i\varepsilon,~~~~i=1,\cdots,\lceil
1/\varepsilon \rceil-1,
\end{equation*}
where $\lceil x \rceil$ is the smallest integer that is greater than
or equal to $x$. Furthermore, denote $t_0=0$ and $t_{\lceil
1/\varepsilon \rceil}=+\infty$. For $i= 1,\cdots,\lceil
1/\varepsilon \rceil$, define brackets $[L_i,U_i]$ with
\begin{equation*}
L_i(s)=-P\left\{\frac{\Delta 1(s\geq \epsilon_{0})1(t_i\geq
\epsilon_{0})}{h^{(1)}(\alpha_0,\epsilon_{0})^2}\right\},
~~U_i(s)=-P\left\{\frac{\Delta 1(s\geq \epsilon_{0})1(t_{i-1}\geq
\epsilon_{0})}{h^{(1)}(\alpha_0,\epsilon_{0})^2}\right\}
\end{equation*}
such that $L_i(s)\leq -P\left\{\frac{\Delta 1(s\geq
\epsilon_{0})1(t\geq
\epsilon_{0})}{h^{(1)}(\alpha_0,\epsilon_{0})^2}\right\}\leq U_i(s)$
when $t_{i-1}< t\leq t_i$. Since
\begin{eqnarray*}
E|U_i-L_i|&\leq& pr(t_{i-1}<\varsigma_{0}\leq
t_i)/\{h^{(1)}(\alpha_0,\tau)\}^2=\varepsilon/\xi^2
\end{eqnarray*}
from Condition 4, we have $
N_{[~]}(\varepsilon/\xi^2,\mathcal{F}_1,L_1)\leq 2/\varepsilon $
which yields
\begin{equation*}
N_{[~]}(\varepsilon,\mathcal{F}_1,L_1)\leq K_1/\varepsilon,
\end{equation*}
where $K_1=2\xi^2$. Similarly, we have
\begin{equation*}
N_{[~]}(\varepsilon,\mathcal{F}_2,L_1)\leq K_2/\varepsilon,
\end{equation*}
where $K_2=2\xi$. From Theorem 2.14.9 in \cite{van and
wellner:1996}, we have
\begin{eqnarray}
&& P^*\bigg(\sup_{t\in [C-E_5,\tau]}
|\mathbb{G}_n\{(1-\eta_0(t))m_1(\alpha_0,\epsilon_0;t)\}|>q\bigg) \nonumber \\
&& \qquad \le P^*\bigg(\sup_{t\in [C-E_5,\tau]}
|\mathbb{G}_n\{m_1(\alpha_0,\epsilon_0;t)\}|>q\bigg)
 \leq
D_1qe^{-2q^2},\label{tailprob1}\\
&& P^*\bigg(\sup_{t\in [C-E_5,\tau]}
|\mathbb{G}_n\{(1-\eta_0(t))m_2(\alpha_0,\epsilon_0;t,\Delta)\}|>q\bigg) \nonumber \\
&& \qquad \le P^*\bigg(\sup_{t\in [C-E_5,\tau]}
|\mathbb{G}_n\{m_2(\alpha_0,\epsilon_0;t,\Delta)\}|>q\bigg) \leq
D_2qe^{-2q^2}\label{tailprob2}
\end{eqnarray}
for some constant $D_1$ depends on $K_1$ and constant $D_2$ depends
on $K_2$. We now show $\sup_{t\in
[C-E_5,\tau]}|\dot{F}_{\alpha}(\alpha_0,t)|$ is bounded.  Direct
calculation yields
\begin{eqnarray*}
&& \sup_{t\in [C-E_5,\tau]}|\dot{F}_{\alpha}(\alpha_0,t)| \\
&& \qquad =\sup_{t\in [C-E_5,\tau]} e^{-\int_{u\leq
t}\frac{dh^{(0)}(\alpha_0,u)}{h^{(1)}(\alpha_0,u)}}\left|\int_{u\leq
t}\frac{d\dot{h}^{(0)}_{\alpha}(\alpha_0,u)}{h^{(1)}(\alpha_0,u)}
-\int_{u\leq
t}\frac{\dot{h}^{(1)}_{\alpha}(\alpha_0,u)dh^{(0)}(\alpha_0,u)}{\{h^{(1)}(\alpha_0,u)\}^2}\right| \\
&&  \qquad \leq \{h^{(1)}(\alpha_0,\tau)\}^{-1}\sup_{t\in
[C-E_5,\tau]}|\dot{h}^{(0)}_{\alpha}(\alpha_0,t)| \\
&&\qquad \qquad  +\{h^{(1)}(\alpha_0,\tau)\}^{-2}
\sup_{u\in(-\infty,\infty)}
\left|\dot{h}_{u}^{(0)}(\alpha_0,u)\right|\sup_{t\in
[C-E_5,\tau]}\int_{u\leq t}|\dot{h}^{(1)}_{\alpha}(\alpha_0,u)|du,
\end{eqnarray*}
where $\dot{h}^{(0)}_{\alpha}(\alpha_0,t)=\frac{\partial}{\partial
\alpha}{h}^{(0)}(\alpha,t)\big|_{\alpha=\alpha_0}$,
$\dot{h}^{(1)}_{\alpha}(\alpha_0,t)=\frac{\partial}{\partial
\alpha}{h}^{(1)}(\alpha,t)\big|_{\alpha=\alpha_0}$ and
$\dot{h}_{u}^{(0)}(\alpha_0,u)=\frac{\partial}{\partial
u}{h}^{(0)}(\alpha_0,u)$. Since
\begin{eqnarray*}
h^{(0)}(\alpha,t)&=&\int\eta_{0}(\min(t+x'\alpha-x'\alpha_0,C-x'\alpha_0))dF_X(x)\\
&=&\int_{x'\alpha\geq
C-t}\eta_{0}(C-x'\alpha_0)dF_X(x)+\int_{x'\alpha<C-t}\eta_{0}(t+x'\alpha-x'\alpha_0)dF_X(x),\\
h^{(1)}(\alpha,t)&=&\int_{x'\alpha\leq
C-t}\{1-\eta_{0}(t+x'\alpha-x'\alpha_0)\}dF_X(x),
\end{eqnarray*}
where $F_X(x)$ is the distribution function of $X$, from Condition 3
we have
\begin{eqnarray*}
&&\sup_{t\in
[C-E_5,\tau]}|\dot{h}^{(0)}_{\alpha}(\alpha_0,t)|=\sup_{t\in
[C-E_5,\tau]}\bigg|\dot{\eta}_{0,\alpha_0}(t)\int_{t+x'\alpha_0<C}xdF_X(x)\bigg|\leq
E_3E|X|<\infty,\\
&&\sup_{u\in(-\infty,\infty)}
\left|\dot{h}_{u}^{(0)}(\alpha_0,u)\right| \le \sup_{u\in(-\infty,\infty)}|\dot{\eta}_{0,\alpha_0}(u)| \leq E_3,\\
 &&\sup_{t\in [C-E_5,\tau]}\int_{u\leq
t}|\dot{h}^{(1)}_{\alpha}(\alpha_0,u)|du\\
&& \qquad \leq \sup_{t\in [C-E_5,\tau]}\int_{u\leq t}
\bigg|\int_{t+x'\alpha_0\leq C}xdF_X(x)\bigg|\dot{\eta}_{0,\alpha_0}(u)du+\int_{-\infty}^{\infty}|x|dF_X(x)\\
&& \qquad \leq E|X|E_3+E|X|<\infty.
\end{eqnarray*}
Since it can be shown that $m_3(\alpha_0,\epsilon_0;X,\Delta)$ has
finite second moment, we have $\sup_{t\in [C-E_5,\tau]}
\mathbb{G}_n[\dot{F}_{\alpha}(\alpha_0,t)m_3(\alpha_0,\epsilon_0;X,\Delta)]
= O_{p^*}(1)$, thus obtain the desired result.
\end{proof}

\begin{lemma} \label{Lemma lod2} Suppose Condition 7 holds, we have
that
\begin{equation}
\left\{\Delta \{Y-\dot{b}(D'(t)\theta)\}D(t), \theta\in\Theta, t \in
{\cal T} \subset \mathbb{R}\right\}\label{lodlemma41}
\end{equation}
 is Donsker.
 \end{lemma}
\begin{proof}
From Condition 7 we know that $\ddot{b}(\cdot)$ is bounded, hence
$\dot{b}(\cdot)$ is a Lipschitiz function. From Theorem 2.10.6 in
\cite{van and wellner:1996}, we know that $D(t)$ and
$\dot{b}(D'(t)\theta)$ are Donsker, hence (\ref{lodlemma41}) is
Donsker.
\end{proof}

\begin{lemma}\label{donskereta} Suppose ${\cal X}$ and ${\cal A}$ be the bounded covariate and parameter spaces. Let ${\cal H}$ be a collection of distribution functions satisfying Condition 3. We
have $\mathcal{F}=\{\eta(t-x'\alpha),t\in{\cal T}\subset \mathbb{R},
x\in{\cal X},\alpha\in \mathcal{A},\eta \in \mathcal{H}\}$ is
Donsker.
\end{lemma}
\begin{proof}
Let $\mathcal{F}_1=\{\eta(t)\}$. From Theorem 2.7.5 in \cite{van and
wellner:1996}, the number of brackets $[L_i,U_i]$ such that
$L_i(t)\leq \eta(t)\leq U_i(t)$ for any nontrivial $\varepsilon$
with $1>\varepsilon>0$ and $\int |U_i(t)-L_i(t)|d\eta_0(t)\leq
\varepsilon$ satisfies $\log
N_{[~]}(\varepsilon,\mathcal{F}_1,L_1(P))\leq K_1/\varepsilon$,
where $K_1<\infty$ is a constant.

For notational simplicity, we consider 1-dimensional $\mathcal{A}$.
Because ${\cal A}$ is bounded, we partition ${\cal A}$ by a set of
intervals  $[l_k,u_k)$ such that $| u_k-l_k | \leq \varepsilon$.
Hence the number of such intervals is bounded by $K_2/\varepsilon$
with a constant $K_2<\infty$. Now we construct brackets for
$\mathcal{F}\equiv\{\eta(t-x\alpha)\}$. Define
$$O_{ik}(t,x)=\min(L_i(t-xu_k),L_i(t-xl_k)),~S_{ik}(t,x)=\max(U_i(t-xu_k),U_i(t-xl_k)).$$
We have
\begin{eqnarray*}
O_{ik}(t,x)&\leq& \min(\eta(t-xl_k),\eta(t-xu_k)) \\
&\leq&
\eta(t-x\alpha) \\
&\leq& \max(\eta(t-xl_k),\eta(t-xu_k))\leq S_{ik}(t,x).
\end{eqnarray*} Since
\begin{eqnarray}
&&P\mid S_{ik}-O_{ik}\mid\nonumber\\
&&\leq \int_{-\infty}^{\infty}\int_{-\infty}^{\infty}\mid U_i(t-xu_k)-L_i(t-xu_k)\mid d\eta_0(t+x\alpha_0)dF_X(x)\label{ul1}\\
&&\hspace{5mm}+\int_{-\infty}^{\infty}\int_{-\infty}^{\infty}\mid U_i(t-xl_k)-L_i(t-xl_k)\mid d\eta_0(t+x\alpha_0)dF_X(x)\label{ul2}\\
&&\hspace{5mm}+\int_{-\infty}^{\infty}\int_{-\infty}^{\infty}\mid U_i(t-xl_k)-L_i(t-xu_k)\mid d\eta_0(t+x\alpha_0)dF_X(x)\label{ul3}\\
&&\hspace{5mm}+\int_{-\infty}^{\infty}\int_{-\infty}^{\infty}\mid
U_i(t-xu_k)-L_i(t-xl_k)\mid d\eta_0(t+x\alpha_0)dF_X(x).\label{ul4}
\end{eqnarray}

Since $[L_i,U_i]$ are brackets for $\mathcal{F}_1$, we have
$(\ref{ul1})\leq \varepsilon$ and $(\ref{ul2})\leq \varepsilon$.
Furthermore, by integration by parts and change of variables we
obtain
\begin{eqnarray*}
(\ref{ul3})&\leq& 2\varepsilon+
\int_0^{\infty}\int_{-\infty}^{\infty}
\{\eta(t-xl_k)-\eta(t-xu_k)\}d\eta_0(t+x\alpha_0)dF_X(x)\\
&&\qquad +\int_{-\infty}^{0}\int_{-\infty}^{\infty}
\{\eta(t-xu_k)-\eta(t-xl_k)\}d\eta_0(t+x\alpha_0)dF_X(x)\\
&=&2\varepsilon+ \int_0^{\infty}\int_{-\infty}^{\infty}
\{\eta_0(t+x\alpha_0 + xu_k)-\eta_0(t+x\alpha_0 + xl_k)\}d\eta(t)dF_X(x)\\
&&\qquad +\int_{-\infty}^{0}\int_{-\infty}^{\infty}
\{\eta_0(t+x\alpha_0+ xl_k)-\eta_0(t+x\alpha_0+xu_k)\}d\eta(t)dF_X(x)\\
&\leq& 2\varepsilon+ \int_0^{\infty}\int_{-\infty}^{\infty}
E_3 x(u_k-l_k)d\eta(t)dF_X(x)  -\int_{-\infty}^{0}\int_{-\infty}^{\infty}E_3 x(u_k-l_k)d\eta(t)dF_X(x)\\
&\leq& 2\varepsilon+E_3 E|X| \varepsilon = K_3 \varepsilon,
\end{eqnarray*}
where $E_3$ is defined in Condition 3, and $K_3=2+E_3 E|X| <
\infty$. Similarly, we have $(\ref{ul4})\leq K_3\varepsilon$. Hence
we have $N_{[~]}((2+2K_3)\varepsilon,\mathcal{F},L_1(P))\leq
\exp(K_1/\varepsilon)K_2/\varepsilon$, i.e.
$N_{[~]}(\varepsilon,\mathcal{F},L_1(P))\leq
\exp(K_1(2+2K_3)/\varepsilon)K_2(2+2K_3)/\varepsilon\leq
\exp((K_1+K_2)(2+2K_3)/\varepsilon)$. Hence, $\mathcal{F}$ is
Donsker.
\end{proof}

\begin{lemma}\label{donsker1}
Suppose Conditions 2, 5-9 hold, we have
\begin{eqnarray}
&&\bigg\{\frac{\int_{C-x'\alpha}^{\tau} f_{\theta,\phi}(y\mid
t+x'\alpha,x) \{y-\dot{b}(D'(t+x'\alpha)\theta)\}D(t+x'\alpha)
d\eta(t)} {\int_{C-x'\alpha}^{\tau} f_{\theta,\phi}(y\mid
t+x'\alpha,x)d\eta(t)}:\label{loddonsker1} \\
&& \qquad \theta\in \Theta, |1/a(\phi)|< l, \alpha \in {\cal A},
\eta \in {\cal H}, \rho\{(\phi,\alpha,\eta),(\phi_0,\alpha_0,\eta_0)
\}<\delta_2, x\in {\cal X}, y \in {\cal Y} \bigg\}\nonumber
\end{eqnarray}
is Donsker.
\end{lemma}
\begin{proof}
From Condition 9, we have
$\{\int_{C-x'\alpha}^{\tau}f_{\theta,\phi}(y\mid
t+x'\alpha,x)d\eta(t)\}$ bounded away from zero. From Section 2.10.2
of \cite{van and wellner:1996}, we only need to show that both the
numerator and denominator in (\ref{loddonsker1}) belong to Donsker
classes. By integration by parts, we have
\begin{eqnarray*}
&& \int_{C-x'\alpha}^{\tau}f_{\theta,\phi}(y \mid t+x'\alpha,x)d\eta(t) \\
&& \qquad =  f_{\theta,\phi}(y\mid \tau+x'\alpha,x)\eta(\tau)-
f_{\theta,\phi}(y\mid C,x)\eta(C-x'\alpha)\\
&&\qquad \qquad - \int_{C-E_5}^{\tau}1(t\geq
C-x'\alpha)\eta(t)f_{\theta,\phi}(y\mid t+x'\alpha,x)\\
&&\qquad \qquad \qquad \gamma
\{y-\dot{b}(D'(t+x'\alpha)\theta)\}\dot{h}(t+x'\alpha)/a(\phi)dt.
\end{eqnarray*}
In the above, $\dot{h}(\cdot)$ is Lipschitz by Condition 2 and
$f_{\theta,\phi}(y\mid t+x'\alpha,x)$ is Lipschitz function for
$\theta$, $\phi$ and $\alpha$ by Conditions 2, 5 and 8, thus both
belong to Donsker classes by Theorem 2.10.6 in \cite{van and
wellner:1996}. By Lemma  \ref{donskereta} we know that
$\{\eta(C-x'\alpha)\}$ is Donsker. Since the class of indicator
functions of half spaces is a VC-class, see e.g. Exercise 9 on page
151 and Exercise 14 on page 152 in \cite{van and wellner:1996}, thus
the set of functions $\{1(t\geq C-x'\alpha)\}$ is a Donsker class.
By Theorem 2.10.3 in \cite{van and wellner:1996}, the permanence of
the Donsker property for the closure of the convex hull,
 we have $\bigg\{\int_{C-E_5}^{\tau}1(t\geq
C-x'\alpha)\eta(t)f_{\theta,\phi}(y\mid t+x'\alpha,x)\gamma
\{y-\dot{b}(D'(t+x'\alpha)\theta)\}/a(\phi)\dot{h}(t+x'\alpha)dt\bigg\}$
is Donsker. Hence the denominator of (\ref{loddonsker1}) belongs to
a Donsker class.

Similarly, by integration by parts,
\begin{eqnarray*}
&&\int_{C-x'\alpha}^{\tau}f_{\theta,\phi}(y\mid t+x'\alpha,x)\{y-\dot{b}(D'(t+x'\alpha)\theta)\}D(t+x'\alpha)d\eta(t)\\
&& \qquad =
f_{\theta,\phi}(y\mid \tau+x'\alpha,x)\{y-\dot{b}(D'(\tau+x'\alpha)\theta)\}D(\tau+x'\alpha)\eta(\tau)\\
&&\qquad -
f_{\theta,\phi}(y\mid C,x)\{y-\dot{b}(D'(C)\theta)\}D(C)\eta(C-x'\alpha)\\
&& \qquad - \int_{C-E_5}^{\tau}1(t\geq
C-x'\alpha)\eta(t)f_{\theta,\phi}(y\mid t+x'\alpha,x)\\
&&\qquad \qquad (\gamma
\{y-\dot{b}(D'(t+x'\alpha)\theta)\}^2D(t+x'\alpha)/a(\phi)-\ddot{b}(D'(t+x'\alpha)\theta)\gamma
D(t+x'\alpha)\\
&&\qquad \qquad
+\{y-\dot{b}(D'(t+x'\alpha)\theta)\}J_{p+2})\dot{h}(t+x'\alpha)dt,
\end{eqnarray*}
where $J_{p+2}=(0,\cdots,0,1)_{1\times (p+2)}'$. Similar to the
denominator, we can show that the above function, which is the
numerator of (\ref{loddonsker1}), belongs to a Donsker class
provided that $\{\ddot{b}(D'(t+x'\alpha)\theta)\}$ is Donsker from
Condition 7.
\end{proof}

\begin{lemma}\label{ANc12} Under Conditions 5-9, when $\theta\rightarrow
\theta_0$ and
$\rho\{(\phi,\alpha,\eta),(\phi_0,\alpha_0,\eta_0)\}\rightarrow 0$,
we have that $
E|\psi_{\theta}(\phi,\alpha,\eta)-\psi_{\theta_0}(\phi_0,\alpha_0,\eta_{0})|^2\rightarrow
0$.
\end{lemma}

\begin{proof} The proof follows straightforward algebraic calculations based on the Mean Value Theorem. The details are thus omitted.
\end{proof}

\begin{lemma}\label{ANc2} Suppose Conditions 2, 5-9 hold, we have
$ E|\psi_{\theta_0}(\phi_0,\alpha_0,\eta_0)|^2<\infty. $
\end{lemma}
\begin{proof}
Again, the proof is based on direct calculation.
\end{proof}


\section{Proof of Theorem 1}
\subsection{Proof of consistency}
\begin{proof}
We prove consistency using Lemma  \ref{Lemma consist}. Since
$\hat{\phi}_n$ and $\hat{\alpha}_n$ are $n^{1/2}$ -consistent, see
the last paragraph of Section 3, and $\hat{\eta}_{n,\hat{\alpha}_n}$
is also $n^{1/2}$ -consistent in a finite interval from Lemma
\ref{rootnconsis}, we have
 \begin{equation*}
\rho\{(\hat{\phi}_n,\hat{\alpha}_n,\hat{\eta}_{n,\hat{\alpha}_n}),(\phi_0,\alpha_0,\eta_{0})\}=o_{p^*}(1).\label{lodthm12}
 \end{equation*}
Given
 that $\theta_0$ is the unique solution to
$\Psi_{\theta}(\phi_0,\alpha_0,\eta_{0})=0$ from Condition 1, we
only need to show that
 \begin{equation}
 \sup_{\theta\in\Theta,\rho\{(\phi,\alpha,\eta),(\phi_0,\alpha_0,\eta_{0})\}\leq
 \delta_n}|\Psi_{\theta,n}(\phi,\alpha,\eta)-\Psi_{\theta}(\phi_0,\alpha_0,\eta_{0})|=o_{p^*}(1)\label{lodthm11}
 \end{equation}
 for every sequence $\delta_n\downarrow 0$. Now

 \begin{eqnarray}
 &&\sup_{\theta\in\Theta,\rho\{(\phi,\alpha,\eta),(\phi_0,\alpha_0,\eta_{0})\}\leq
 \delta_n}|\Psi_{\theta,n}(\phi,\alpha,\eta)-\Psi_{\theta}(\phi_0,\alpha_0,\eta_{0})| \nonumber \\
 && \qquad \leq \sup_{\theta\in\Theta}|({\mathbb P}_n-P)[\Delta \{Y-\dot{b}(D(T)\theta)\}D(T)]| \label{lodthm131}\\
 && \qquad \quad +\sup_{\theta\in\Theta,\rho\{(\phi,\alpha,\eta),(\phi_0,\alpha_0,\eta_{0})\}\leq
 \delta_n}\label{lodthm132}\\
&& \qquad \quad P \Bigg|\frac{\int_{C-X'\alpha}^{\tau}
f_{\theta,\phi}(Y\mid t+X'\alpha,X)
\{Y-\dot{b}(D'(t+X'\alpha)\theta)\}D(t+X'\alpha) d\eta(t)}
{\int_{C-X'\alpha}^{\tau}
f_{\theta,\phi}(Y\mid t+X'\alpha,X)d\eta(t)}\nonumber\\
&& \qquad \quad -\frac{\int_{C-X'\alpha_0}^{\tau}
f_{\theta,\phi_0}(Y\mid t+X'\alpha_0,X)
\{Y-\dot{b}(D'(t+X'\alpha_0)\theta)\}D(t+X'\alpha_0) d\eta_{0}(t)}
{\int_{C-X'\alpha_0}^{\tau}
f_{\theta,\phi_0}(Y\mid t+X_i'\alpha_0,X)d\eta_{0}(t)}\Bigg|\nonumber\\
&& \qquad \quad
+\sup_{\theta\in\Theta,\rho\{(\phi,\alpha,\eta),(\phi_0,\alpha_0,\eta_{0})\}\leq
 \delta_n} \bigg|({\mathbb P}_n-P)(1-\Delta)\label{lodthm133}\\
&&\qquad \quad \frac{\int_{C-X'\alpha}^{\tau} f_{\theta,\phi}(Y\mid
t+X'\alpha,X) \{Y-\dot{b}(D'(t+X'\alpha)\theta)\}D(t+X'\alpha)
d\eta(t)}{\int_{C-X'\alpha}^{\tau} f_{\theta,\phi}(Y\mid
t+X'\alpha,X)d\eta(t)}\bigg| =o_{p^*}(1), \nonumber
 \end{eqnarray}
where (\ref{lodthm131}) and (\ref{lodthm133}) equal to $o_{p^*}(1)$
are from Lemma  \ref{Lemma lod2} and Lemma \ref{donsker1},
respectively, and (\ref{lodthm132}) equal to $o_{p^*}(1)$ follows a
direct calculation similar to Lemma \ref{ANc12} using the Mean Value
Theorem.
\end{proof}

\subsection{Proof of asymptotic normality}
\begin{proof}
We now verify all the conditions in Lemma \ref{Lemma an}. Condition
(i) holds because $\{\psi_\theta(\phi, \alpha, \eta)\}$  is Donsker
by Lemmas  \ref{Lemma lod2} and \ref{donsker1}, together with the
result in Lemma \ref{ANc12}. Condition (ii) holds by the classical
central limit theorem for independent and identically distributed
data with
$E|\psi_{\theta_0}(\phi_0,\alpha_0,\eta_0(\alpha_0))|^2<\infty$ from
Lemma \ref{ANc2}.

For Condition (iii), given that
$\rho\{(\hat{\phi}_n,\hat{\alpha}_n,\hat{\eta}_{n,\hat{\alpha}_n}),(\phi_0,\alpha_0,\eta_{0})\}=O_{p^*}(n^{-1/2})$,
taking the Taylor expansion for $\theta$, $\phi$ and $\alpha$ we
obtain
\begin{eqnarray*}
&&\Psi_{\hat{\theta}_n}(\hat{\phi}_n,\hat{\alpha}_n,\hat{\eta}_{n,\hat{\alpha}_n})
-\Psi_{\theta_0}(\phi_0,\alpha_0,\eta_{0})\nonumber\\
&& \qquad
=\dot{\Psi}_{1,\tilde{\theta}}(\hat{\phi}_n,\hat{\alpha}_n,\hat{\eta}_{n,\hat{\alpha}_n})
(\hat{\theta}_n-\theta_0)-\dot{\Psi}_{2,\theta_0}(\tilde{\phi},\hat{\alpha}_n,\hat{\eta}_{n,\hat{\alpha}_n})(\hat{\phi}_n-\phi_0)\nonumber\\
&&\qquad \qquad
-\dot{\Psi}_{3,\theta_0}(\phi_0,\tilde{\alpha},\eta_{0})(\hat{\alpha}_n-\alpha_0)-R(\theta_0,\phi_0,\hat{\alpha}_n,\hat{\eta}_{n,\hat{\alpha}_n},\eta_{0}),
\end{eqnarray*}
where $\tilde{\theta}$ is between $\theta_0$ and $\hat{\theta}_n$,
$\tilde{\phi}$ is between $\phi_0$ and $\hat{\phi}_n$,
$\tilde{\alpha}$ is between $\alpha_0$ and $\hat{\alpha}_n$, and the
remainder has the following form
\begin{eqnarray*}
&&R(\theta_0,\phi_0,\alpha,\eta,\eta_{0})\\
&& \qquad = P\bigg[(1-\Delta)\bigg\{\frac{\int_{C-X'\alpha}^{\tau}
A(t,\theta_0,\phi_0,\alpha)d\eta(t)}{\int_{C-X'\alpha}^{\tau}
B(t,\theta_0,\phi_0,\alpha)d\eta(t)}-\frac{\int_{C-X'\alpha}^{\tau}
A(t,\theta_0,\phi_0,\alpha)d\eta_{0}(t)}{\int_{C-X'\alpha}^{\tau}
B(t,\theta_0,\phi_0,\alpha)d\eta_{0}(t)}\bigg\}\bigg]
\end{eqnarray*}
with
\begin{eqnarray*}
&&A(t,\theta_0,\phi_0,\alpha)=f_{\theta_0,\phi_0}(Y\mid
t+X'\alpha,X)
\{Y-\dot{b}(D'(t+X'\alpha)\theta_0)\}D(t+X'\alpha),\\
&&B(t,\theta_0,\phi_0,\alpha)=f_{\theta_0,\phi_0}(Y\mid
t+X'\alpha,X).
\end{eqnarray*}
It can be show by direct calculation that
$|\dot{\Psi}_{1,\tilde{\theta}}(\hat{\phi}_n,\hat{\alpha}_n,\hat{\eta}_{n,\hat{\alpha}_n})-
\dot{\Psi}_{1,\theta_0}(\phi_0,\alpha_0,\eta_{0})|=o_{p^*}(1)$,
$|\dot{\Psi}_{2,\theta_0}(\tilde{\phi},\hat{\alpha}_n,\hat{\eta}_{n,\hat{\alpha}_n})-
\dot{\Psi}_{2,\theta_0}(\phi_0,\alpha_0,\eta_{0})|=o_{p^*}(1)$ and
$|\dot{\Psi}_{3,\theta_0}(\phi_0,\tilde{\alpha},\eta_{0})-
\dot{\Psi}_{3,\theta_0}(\phi_0,\alpha_0,\eta_{0})|=o_{p^*}(1)$.

Define
\begin{eqnarray}
&&\dot{\Psi}_{4,\theta_0}(\phi_0,\alpha_0,\eta_{0})(\hat{\eta}_{n,\hat{\alpha}_n}-\eta_{0})\label{psi4}\\
&& \qquad = P\bigg[(1-\Delta)\bigg\{\frac{\int_{C-X'\alpha_0}^{\tau}
A(t,\theta_0,\phi_0,\alpha_0)d[\hat{\eta}_{n,\hat{\alpha}_n}(t)-\eta_0(t)]}{\int_{C-X'\alpha_0}^{\tau}
B(t,\theta_0,\phi_0,\alpha_0)d\eta_{0}(t)}\nonumber\\
&&\qquad \qquad -\frac{\int_{C-X'\alpha_0}^{\tau}
A(t,\theta_0,\phi_0,\alpha_0)d\eta_{0}(t)\int_{C-X'\alpha_0}^{\tau}
B(t,\theta_0,\phi_0,\alpha_0)d[\hat{\eta}_{n,\hat{\alpha}_n}(t)-\eta_0(t)]}{\int_{C-X'\alpha_0}^{\tau}
B(t,\theta_0,\phi_0,\alpha_0)d\eta_{0}(t)^2}\bigg\}\bigg]\nonumber\\
&& \qquad = P\bigg[(1-\Delta)\bigg\{\frac{\int_{C-X'\alpha_0}^{\tau}
A(t,\theta_0,\phi_0,\alpha_0)d\hat{\eta}_{n,\hat{\alpha}_n}(t)}{\int_{C-X'\alpha_0}^{\tau}
B(t,\theta_0,\phi_0,\alpha_0)d\eta_{0}(t)}\nonumber\\
&&\qquad \qquad -\frac{\int_{C-X'\alpha_0}^{\tau}
A(t,\theta_0,\phi_0,\alpha_0)d\eta_{0}(t)\int_{C-X'\alpha_0}^{\tau}
B(t,\theta_0,\phi_0,\alpha_0)d\hat{\eta}_{n,\hat{\alpha}_n}(t)}{\int_{C-X'\alpha_0}^{\tau}
B(t,\theta_0,\phi_0,\alpha_0)d\eta_{0}(t)^2}\bigg\}\bigg].\nonumber
\end{eqnarray}
Then we have
\begin{eqnarray*}
&&|R(\theta_0,\phi_0,\hat{\alpha}_n,\hat{\eta}_{n,\hat{\alpha}_n},\eta_{0})-\dot{\Psi}_{4,\theta_0}(\phi_0,\alpha_0,\eta_{0})
(\hat{\eta}_{n,\hat{\alpha}_n}-\eta_{0})|\\
&&\qquad \leq
|R(\theta_0,\phi_0,\hat{\alpha}_n,\hat{\eta}_{n,\hat{\alpha}_n},\eta_{0})-
R(\theta_0,\phi_0,\alpha_0,\hat{\eta}_{n,\hat{\alpha}_n},\eta_{0})|\\
&&\qquad \qquad
+|R(\theta_0,\phi_0,\alpha_0,\hat{\eta}_{n,\hat{\alpha}_n},\eta_{0})-\dot{\Psi}_{4,\theta_0}(\phi_0,\alpha_0,\eta_{0})
(\hat{\eta}_{n,\hat{\alpha}_n}-\eta_{0})|\\
&&\qquad =D_1+D_2.
\end{eqnarray*}
Now
$D_1=o(|\hat{\alpha}_n-\alpha_0|+\|\hat{\eta}_{n,\hat{\alpha}_n}-\eta_{0}\|)$
can be shown by
\begin{eqnarray*}
&&\frac{A_1}{B_1}-\frac{A_2}{B_2}-\frac{A_3}{B_3}+\frac{A_4}{B_4}\\
&& \qquad =\frac{A_1}{B_1B_2}(B_2-B_1-B_4+B_3)+\frac{A_1}{B_1B_2B_3B_4}(B_3B_4-B_1B_2)(B_4-B_3)\\
&&\qquad \qquad
+\frac{A_1-A_3}{B_3B_4}(B_4-B_3)+\frac{A_1-A_2}{B_2B_4}(B_4-B_2)+\frac{A_1-A_2-A_3+A_4}{B_4},
\end{eqnarray*}
and
$D_2=o(|\hat{\alpha}_n-\alpha_0|+\|\hat{\eta}_{n,\hat{\alpha}_n}-\eta_{0}\|)$
can be shown by
\begin{eqnarray*}
&&\frac{A_1}{B_1} - \frac{A_2}{B_2} - \frac{A_1}{B_2} +
\frac{A_2B_1}{B_2^2} =\frac{1}{B_1B_2^2} \{A_1(B_1 - B_2)^2 -
B_1(A_2-A_1)(B_2-B_1)\}.
\end{eqnarray*}

Since $\hat{\phi}_n$, $\hat{\alpha}_n$ and $\hat{\eta}_n$ are all
root-$n$ consistent, under Conditions (i)-(iii), Condition (iv)
holds automatically. Then by Lemma \ref{Lemma an} we have that
$\hat{\theta}_n$ is $n^{1/2}$ -consistent and (\ref{lemma23}) holds
with
\begin{eqnarray*}
&&\dot{\Psi}_{1,\theta_0}(\phi_0,\alpha_0,\eta_{0})\label{A}\\
&&\qquad =E[\Delta
\ddot{b}\{D'(T)\theta_0\}D(T)D'(T)]\nonumber\\
&&\qquad \qquad
-E\bigg[(1-\Delta)\bigg\{\int_{C-X'\alpha_0}^{\tau}f_{\theta_0,\phi_0}(Y\mid
t+X'\alpha_0,X)
d\eta_{0}(t)\bigg\}^{-2}\nonumber\\
&&\qquad \qquad \qquad
\bigg(\int_{C-X'\alpha_0}^{\tau}f_{\theta_0,\phi_0}(Y\mid
t+X'\alpha_0,X)\{Y-\dot{b}(D'(t+X'\alpha_0)\theta_0)\} \nonumber \\
&& \qquad \qquad \qquad D(t+X'\alpha_0) d\eta_{0}(t)\bigg)^{\otimes
2} \bigg]\nonumber,
\end{eqnarray*}

\begin{eqnarray*}
&&\dot{\Psi}_{2,\theta_0}(\phi_0,\alpha_0,\eta_{0})\label{psi2}\\
&& \qquad =-E\bigg[(1-\Delta)\bigg\{\int_{C-X'\alpha_0}^{\tau}f_{\theta_0,\phi_0}(Y\mid t+X'\alpha_0,X)d\eta_{0}(t)\bigg\}^{-1}\nonumber\\
&&\qquad \qquad
\bigg\{\int_{C-X'\alpha_0}^{\tau}f_{\theta_0,\phi_0}(Y\mid
t+X'\alpha_0,X)\{Y-\dot{b}(D'(t+X'\alpha_0)\theta_0)\}D(t+X'\alpha_0)\nonumber\\
&&\qquad \qquad
\Big([Y \{D'(t+X'\alpha_0)\theta_0\}-b(D'(t+X'\alpha_0)\theta_0)]a'(\phi_0)/a(\phi_0)^2-\dot{c}_{\phi}(Y,\phi_0)\Big)d\eta_{0}(t)\bigg\}\nonumber\\
&&\qquad \qquad
-\bigg\{\int_{C-X'\alpha_0}^{\tau}f_{\theta_0,\phi_0}(Y\mid
t+X'\alpha_0,X)d\eta_{0}(t)\bigg\}^{-2}
\bigg\{\int_{C-X'\alpha_0}^{\tau}f_{\theta_0,\phi_0}(Y\mid t+X'\alpha_0,X)\nonumber\\
&&\qquad \qquad \Big([Y
\{D'(t+X'\alpha_0)\theta_0\}-b(D'(t+X'\alpha_0)\theta_0)]a'(\phi_0)/a(\phi_0)^2-\dot{c}_{\phi}(Y,\phi_0)\Big)d\eta_{0}(t)\nonumber\\
&&\qquad \qquad \int_{C-X'\alpha_0}^{\tau}f_{\theta_0,\phi_0}(Y\mid
t+X'\alpha_0,X)\{Y-\dot{b}(D'(t+X'\alpha_0)\theta_0)\}D(t+X'\alpha_0)
d\eta_{0}(t)\bigg\}\bigg], \nonumber
\end{eqnarray*}
and
\begin{eqnarray*}
&&\dot{\Psi}_{3,\theta_0}(\phi_0,\alpha_0,\eta_{0})\label{psi3}\\
&& \qquad =-E\bigg[(1-\Delta)\bigg\{\int_{C-X'\alpha_0}^{\tau}f_{\theta_0,\phi_0}(Y\mid t+X'\alpha_0,X)d\eta_{0}(t)\bigg\}^{-2}\nonumber\\
&& \qquad \qquad \int_{C-X'\alpha_0}^{\tau}f_{\theta_0,\phi_0}(Y\mid
t+X'\alpha_0,X)\{Y-\dot{b}(D'(t+X'\alpha_0)\theta_0)\}D(t+X'\alpha_0)
d\eta_{0}(t) \nonumber\\
&&\qquad \qquad
\bigg\{\int_{C-X'\alpha_0}^{\tau}f_{\theta_0,\phi_0}(Y\mid
t+X'\alpha_0,X)\{Y-\dot{b}(D'(t+X'\alpha_0)\theta_0)\} \nonumber \\
&& \qquad \qquad \gamma_0 X'
\dot{h}(t+X'\alpha_0)/a(\phi_0)d\eta_{0}(t)+f_{\theta_0,\phi_0}(Y\mid
C,X) \dot{\eta}_{0}(C-X'\alpha_0)X'\bigg\}\bigg]\nonumber.
\end{eqnarray*}

Finally, we obtain
\begin{eqnarray}\label{Thman1}
&&n^{1/2}\{(\Psi_{n,\theta_0}-\Psi_{\theta_0})(\phi_0,\alpha_0,\eta_{0})\}
=\mathbb{G}_n\bigg(\Delta
\{Y-\dot{b}(D'(T)\theta_0)\}D(T)\nonumber\\
&&\qquad+(1-\Delta)\bigg\{\int_{C-X'\alpha_0}^{\tau}
f_{\theta_0,\phi_0}(Y\mid t+X'\alpha_0,X)d\eta_{0}(t)\bigg\}^{-1}\nonumber\\
&&\qquad \int_{C-X'\alpha_0}^{\tau} f_{\theta_0,\phi_0}(Y\mid
t+X'\alpha_0,X)
\{Y-\dot{b}(D'(t+X'\alpha_0)\theta_0)\}D(t+X'\alpha_0)
d\eta_{0}(t)\bigg)\nonumber\\
&&
=\mathbb{G}_n\bigg\{G_1(\theta_0,\phi_0,\alpha_0,\eta_0,\Delta,Y,X,V)\bigg\}
\end{eqnarray}

and
\begin{eqnarray}
&&
n^{1/2}\dot{\Psi}_{2,\theta_0}(\phi_0,\alpha_0,\eta_{0})(\hat{\phi}_n-\phi_0)\}
\label{Thman2} \\
&& \qquad
=\mathbb{G}_n\bigg\{\dot{\Psi}_{2,\theta_0}(\phi_0,\alpha_0,\eta_{0})m_4(\theta_0,\Delta,Y,X,V)\bigg\}+o_{p}(1),
\nonumber
\end{eqnarray}
where
$n^{1/2}(\hat{\phi}_n-\phi_0)=\mathbb{G}_n{m_4(\theta_0,\phi_0,Y,X)}+o_p(1)$
with $m_4(\theta_0,\phi_0,Y,X)=\Delta\{Y-D'(T)\theta_0\}^2$ for
linear regression and $m_4=0$ for the logistic and Poisson
regressions. For Gehan weighted estimate $\hat{\alpha}_n$, we have
\begin{eqnarray}
n^{1/2}\dot{\Psi}_{3,\theta_0}(\phi_0,\alpha_0,\eta_{0})
(\hat{\alpha}_n-\alpha_0)=\mathbb{G}_n\bigg\{\dot{\Psi}_{3,\theta_0}(\phi_0,\alpha_0,\eta_{0})
m_3(\alpha_0,\epsilon_0;\Delta,X)\bigg\}+o_{p}(1).\label{Thman3}
\end{eqnarray}
Furthermore, from (\ref{eta}) and (\ref{psi4}) we obtain
\begin{eqnarray}\label{Thman4}
&&n^{1/2}\dot{\Psi}_{4,\theta_0}(\phi_0,\alpha_0,\eta_{0})(\hat{\eta}_{n,\hat{\alpha}_n}-\eta_{0})\\\nonumber
&& =
\mathbb{G}_n\bigg[-\int_{\mathcal{X}}\int_{-\infty}^{\infty}(1-\Delta)\bigg\{f_{\theta_0,\phi_0}(y\mid
\tau+x'\alpha_0,x)\bigg(\{1-\eta_{0}(\tau)\}\{m_1(\alpha_0,\epsilon_0;\tau)\\\nonumber
&&\qquad
+m_2(\alpha_0,\epsilon_0;\tau,\Delta)\}+\dot{F}_{\alpha}(\alpha,\tau)m_3(\alpha_0,\epsilon_0;x,\Delta)\bigg)\\\nonumber
&&\quad-f_{\theta_0,\phi_0}(y\mid
C,x)\bigg(\{1-\eta_{0}(C-x'\alpha_0)\}\{m_1(\alpha_0,\epsilon_0;C-x'\alpha_0)\\\nonumber
&&\qquad+m_2(\alpha_0,\epsilon_0;C-x'\alpha_0,\Delta)\}+\dot{F}_{\alpha}(\alpha,C-x'\alpha_0)m_3(\alpha_0,\epsilon_0;x,\Delta)\bigg)\\\nonumber
&&\quad-\int_{C-x'\alpha_0}^{\tau}f_{\theta_0,\phi_0}(y\mid
t+x'\alpha_0,x)\gamma_0
\dot{h}(t+x'\alpha_0)\{y-\dot{b}(D'(t+x'\alpha_0)\theta_0)\}/a(\phi_0)
\\\nonumber
&&\qquad\bigg(\{1-\eta_{0}(t)\}\{m_1(\alpha_0,\epsilon_0;t)+m_2(\alpha_0,\epsilon_0;t,\Delta)\}+\dot{F}_{\alpha}(\alpha,t)m_3(\alpha_0,\epsilon_0;x,\Delta)\bigg)dt\bigg\}\\\nonumber
&&\qquad\bigg(\int_{C-x'\alpha_0}^{\tau} f_{\theta_0,\phi_0}(y\mid
t+x'\alpha_0,x)\{y-\dot{b}(D'(t+x'\alpha_0)\theta_0)\}D(t+x'\alpha_0)d\eta_{0}(t)\bigg)\\\nonumber
&&\qquad \bigg(\int_{C-x'\alpha_0}^{\tau} f_{\theta_0,\phi_0}(y\mid
t+x'\alpha_0,x)d\eta_{0}(t)\bigg)^{-2}dydF_X(x)\bigg]+o_{p}(1)\\
&&=\mathbb{G}_n\{G_2(\theta_0,\phi_0,\alpha_0,\eta_0,\Delta,Y,X,V)\}\nonumber
\end{eqnarray}

Hence,
$(\Psi_{n,\theta_0}-\Psi_{\theta_0})(\phi_0,\alpha_0,\eta_{0})+\dot{\Psi}_{2,\theta_0}(\phi_0,\alpha_0,\eta_{0})
(\hat{\phi}_n-\phi_0)
+\dot{\Psi}_{3,\theta_0}(\phi_0,\alpha_0,\eta_{0})
(\hat{\alpha}_n-\alpha_0)+\dot{\Psi}_{4,\theta_0}(\phi_0,\alpha_0,\eta_{0})
(\hat{\eta}_{n,\hat{\alpha}_n}-\eta_{0})$ is the sum of independent
and identically distributed terms and the classical central limit
theorem applies. We have $\sqrt{n}(\hat{\theta}_n-\theta_0)$
converges weakly to a mean zero normal random variable with variance
$A^{-1}BA^{-1}$, where
\begin{eqnarray*}
&&A=-\dot{\Psi}_{1,\theta_0}(\phi_0,\alpha_0,\eta_0),\\
&&B=\bigg\{G_1(\theta_0,\phi_0,\alpha_0,\eta_0,\Delta,Y,X,V)+\dot{\Psi}_{2,\theta_0}(\phi_0,\alpha_0,\eta_{0})m_4(\theta_0,\Delta,Y,X,V)\\
&&\qquad+\dot{\Psi}_{3,\theta_0}(\phi_0,\alpha_0,\eta_{0})
m_3(\alpha_0,\epsilon_0;\Delta,X)+G_2(\theta_0,\phi_0,\alpha_0,\eta_0,\Delta,Y,X,V)\bigg\}^{\otimes
2}.
\end{eqnarray*}

Note that for other rank based estimates of $\alpha$, $m_3$ in $B$
is the corresponding influence function with different forms; For
the sieve maximum likelihood estimates \citep{ding:2011}, $m_3$ is
the efficient influence function \citep{ritov and wellner:1988}. It
is clearly seen that the analytic form of the asymptotic variance is
too complicated to be useful for the asymptotic variance estimation.
Hence in our numerical studies we use bootstrap to obtain the
variance estimator.
\end{proof}